\documentclass[sigconf,nonacm]{acmart} 
\usepackage{popets}
\usepackage{xcolor}
\usepackage{xspace}

\usepackage{booktabs}
\usepackage{siunitx}
\usepackage{l3draw}
\usepackage{subfig}

\usepackage{tikz}

\ExplSyntaxOn
\NewDocumentCommand { \FilledCircle } { m } {
    \draw_begin:
        \draw_baseline:n { -.1em }
        \draw_path_moveto:n { 0em , 0.75em }
        \draw_path_arc:nnn { 90 } { 90 - #1 * 360 } { 0.5em }
        \draw_path_lineto:n { 0em , 0.25em }
        \draw_path_close:
        \draw_path_use_clear:n { fill }
        \draw_path_circle:nn { 0em , 0.25em } { 0.5em }
        \draw_path_use_clear:n { stroke }
    \draw_end:
}
\ExplSyntaxOff

\usepackage{todonotes}
\setuptodonotes{inline}

\usepackage{acronym}

\acrodef{VM}[VM]{virtual machine}
\acrodef{SME}[SME]{Secure Memory Encryption}
\acrodef{SEV}[SEV]{Secure Encrypted Virtualization}
\acrodef{SNP}[SEV-SNP]{Secure Encrypted Virtualization - Secure Nested Paging}
\acrodef{ES}[SEV-ES]{SEV Encrypted State}
\acrodef{SP}[AMD SEV-SP]{AMD Secure Processor}
\acrodef{SGX}[SGX]{Software Guard Extensions}
\acrodef{TDX}[TDX]{Trust Domain Extensions}
\acrodef{CCA}[CCA]{Confidential Compute Architecture}
\acrodef{TEE}[TEE]{Trusted Execution Environment}
\acrodef{IC}[IC]{Internet Computer}
\acrodef{TCB}[TCB]{Trusted Computing Base}
\acrodef{VCEK}[VCEK]{Versioned Chip Endorsement Key}
\acrodef{VLEK}[VLEK]{Versioned Loaded Endorsement Key}
\acrodef{ASK}[ASK]{AMD SEV Key}
\acrodef{ARK}[ARK]{AMD Root Key}
\acrodef{TLS}[TLS]{Transport Layer Security}
\acrodef{SSL}[SSL]{Secure Socket Layer}
\acrodef{CA}[CA]{Certificate Authority}
\acrodef{HRoT}[HRoT]{Hardware Root of Trust}
\acrodef{IMA}[IMA]{Integrity Measurement Architecture}
\acrodef{CA}[CA]{Certificate Authority}
\acrodefplural{CA}[CAs]{Certificate Authorities}
\acrodef{CT}[CT]{Certificate Transparency}
\acrodef{ACME}[ACME]{Automatic Certificate Management Environment}
\acrodef{CVM}[CVM]{Confidential Virtual Machine}
\acrodef{INRA}[]{Indirect Remote Attestation}
\acrodef{DNS}[DNS]{Domain Name System}
\acrodef{HCL}[HCL]{Hardware Compatibility Layer}
\acrodef{MAA}[MAA]{Microsoft Azure Attestation}
\acrodef{PCR}[PCR]{Platform Configuration Register}
\acrodef{AMI}[AMI]{Amazon Machine Image}
\acrodef{CSP}[CSP]{Cloud Service Provider}
\acrodef{AK}[AK]{Attestation Key}
\acrodef{EK}[EK]{Endorsement Key}
\acrodef{KDS}[KDS]{Key Distribution System}
\acrodef{IaaS}[IaaS]{Infrastructure as a Service}
\acrodef{AWS}[AWS]{Amazon Web Services}
\acrodef{ARM}[ARM]{Azure Resource Manager}
\acrodef{THIM}[THIM]{Trusted Hardware Identity Management}
\acrodef{GCP}[GCP]{Google Cloud Platform}
\acrodef{QE}[QE]{Quoting Enclave}
\acrodef{TPM}[TPM]{Trusted Platform Module}
\acrodef{TD}[TD]{Trust Domain}
\acrodef{CCE}[CCE]{Confidential Computing Enforcement}
\acrodef{NVRAM}[NVRAM]{non-volatile random-access memory}
\acrodef{RATS}[RATS]{Remote Attestation procedures}
\acrodef{PCK}[PCK]{Provisioning Certification Key}
\acrodef{PCE}[PCE]{Provisioning Certification Enclave}
\acrodef{AIK}[AIK]{Attestation Identity Key}
\acrodef{SRK}[SRK]{Storage Root Key}
\acrodef{MRTD}[MRTD]{Measurement of Trust Domain}
\acrodef{RTMR}[RTMR]{Run-Time Measurement Register}
\acrodef{TDQE}[TDQE]{TD Quoting Enclave}
\acrodef{MAC}[MAC]{Message Authentication Code}
\acrodef{VMPL}[VMPL]{Virtual Machine Privilege Level}
\acrodef{DCAP}[DCAP]{Datacenter Attestation Primitives}
\acrodef{OCI}[OCI]{Open Container Initiative}
\acrodef{UVM}[UVM]{Utility VM}
\acrodef{TME-MK}[TME-MK]{Total Memory Encryption – Multi-Key}
\acrodef{JWKS}[JWKS]{JSON Web Key Set}

\setcopyright{popets}
\copyrightyear{YYYY}

\acmYear{YYYY}
\acmVolume{YYYY}
\acmNumber{X}
\acmDOI{XXXXXXX.XXXXXXX}
\acmISBN{}
\acmConference{Proceedings on Privacy Enhancing Technologies}
\usepackage{graphicx} 

\title{SoK: A cloudy view on trust relationships of CVMs}
\subtitle{How Confidential Virtual Machines are falling short in Public Cloud}

\date{February 2025}

\keywords{Confidential Computing, Remote Attestation, AMD SEV-SNP, Intel TDX, Intel SGX, Cloud Security, Systematisation of Knowledge}

\begin{document}

\settopmatter{printacmref=false} 
\renewcommand\footnotetextcopyrightpermission[1]{} 

\author{Jana Eisoldt}
\orcid{0009-0003-1729-7963}
\affiliation{%
  \institution{Barkhausen Institut}
  \city{Dresden}
  \country{Germany}}
\email{jana.eisoldt@barkhauseninstitut.org}

\author{Anna Galanou}
\orcid{0002-4148-7631}
\affiliation{%
  \institution{TU Dresden}
  \country{Germany}
}
  \email{anna.galanou@tu-dresden.de}

\author{Andrey Ruzhanskiy}
\affiliation{%
  \institution{Deutsche Telekom MMS GmbH}
  \country{Germany}
}
\email{andrey.ruzhanskiy@telekom.de}

\author{Nils Küchenmeister}
\orcid{0009-0004-0376-0328}
\affiliation{%
  \institution{Deutsche Telekom MMS GmbH}
  \country{Germany}
}
\email{nils.kuechenmeister@tu-dresden.de}

\author{Yewgenij Baburkin}
\orcid{0009-0007-0029-3877}
\affiliation{
    \institution{Deutsche Telekom MMS GmbH}
    \country{Germany}
}
\email{yewgenij.baburkin@telekom.de}

\author{Tianxiang Dai}
\orcid{0009-0002-7968-2499}
\affiliation{
  \institution{Lancaster University Leipzig}
  \country{Germany}
}
\email{t.dai@lancaster.ac.uk}

\author{Ivan Gudymenko}
\affiliation{
\institution{Deutsche Telekom MMS GmbH}
\country{Germany}
}
\email{ivan.gudymenko@telekom.de}

\author{Stefan Köpsell}
\affiliation{%
  \institution{Barkhausen Institut}
  \city{Dresden}
  \country{Germany}
}
\email{stefan.koepsell@barkhauseninstitut.org}
  
\author{Rüdiger Kapitza}
\affiliation{
\institution{FAU Erlangen-Nürnberg}
\country{Germany}
}
\email{ruediger.kapitza@fau.de}

\begin{abstract}
Confidential computing in the public cloud intends to safeguard workload privacy while outsourcing infrastructure management to a cloud provider. 
This is achieved by executing customer workloads within so called \aclp{TEE}, such as \acp{CVM}, which protect them from unauthorized access by cloud administrators and privileged system software. 
At the core of confidential computing lies remote attestation—a mechanism that enables workload owners to verify the initial state of their workload and furthermore authenticate the underlying hardware.  

While this represents a significant advancement in cloud security, this SoK critically examines the confidential computing offerings of market-leading cloud providers to assess whether they genuinely adhere to its core principles.
We develop a taxonomy based on carefully selected criteria to systematically evaluate these offerings, enabling us to analyse the components responsible for remote attestation, the evidence provided at each stage, the extent of cloud provider influence and whether this undermines the threat model of confidential computing.
Specifically, we investigate how \acp{CVM} are deployed in the public cloud infrastructures, the extent to which customers can request and verify attestation evidence, and their ability to define and enforce configuration and attestation requirements. 
This analysis provides insight into whether confidential computing guarantees—namely confidentiality and integrity—are genuinely upheld. 
Our findings reveal that all major cloud providers retain control over critical parts of the trusted software stack and, in some cases, intervene in the standard remote attestation process. 
This directly contradicts their claims of delivering confidential computing, as the model fundamentally excludes the cloud provider from the set of trusted entities.
\end{abstract}

\acresetall

\maketitle

\section{Introduction}
\label{sec:int}

Confidential computing has gained significant traction in public cloud environments \cite{GoogleConfComp, azureConfComp, awsConfComp}, promising privacy and enhanced security for sensitive workloads by migrating them into hardware-protected \acp{TEE}. 
These guarantee protection against unauthorised access from external entities, including threats from the cloud providers themselves. 
Early research systems that paved the way for this development, such as Haven~\cite{baumann15haven}, Scone~\cite{arnautov16scone}, and Graphene-SGX~\cite{tsai17graphene-sgx}, envisioned a system model in which users would not lose control of their code and data while using cloud resources, but their integrity and confidentiality would be maintained.
This is achieved by enforcing a clear separation of trust:
Cloud customers would control their workloads, while untrusted infrastructure—including the host operating system, hypervisor, and cloud provider—would only be trusted to provide and manage resources.
This leaves the hardware vendor of the used servers entrusted with the actual integrity and confidentiality protection provided by the hardware mechanisms and associated \ac{TCB} (e.g., firmware) required to establish the \acp{TEE}.
In fact, these assumptions still build the basis for recent research systems~\cite{johnson2024confidential, ge2022hecate} because they are based on the fundamental concepts of trusted execution and are backed by current hardware extensions, such as Intel \ac{SGX}~\cite{sgxreference}, AMD \ac{SNP}~\cite{AMDSEVSNP} and Intel \ac{TDX}~\cite{IntelTDXDocumentation}. 
By providing strong hardware-backed isolation, coupled with remote attestation, these hardware extensions allow users to verify the integrity and protection of their execution environments without relying on infrastructure providers.

However, as confidential computing evolved and is increasingly embraced by public cloud providers who aim to address customer needs, security concerns, and regulatory requirements, the described idealised separation of trust has been eroded by a transfer of control over the \ac{TCB} enabling \acp{TEE}.
Cloud providers began to integrate their own middleware and attestation mechanisms, effectively acting as gatekeepers for verifying the \ac{TEE} integrity and authenticity. 
This is particularly concerning, since remote attestation is the foundation of confidential computing, enabling \ac{TEE} owners/users to verify the integrity and authenticity of their execution environment before handling sensitive data. 
Ideally, attestation should provide cryptographic proof that the workload runs in an untampered, hardware-enforced \ac{TEE}, isolated from any external interference. 
However, the actual trust models underpinning remote attestation in public cloud environments often diverge from the assumptions users might expect. 
More specifically, even though the typical \ac{TEE} threat model considers the cloud provider untrusted, most of the current cloud offerings still require users to rely on the provider itself to perform attestation. 
This creates a fundamental mismatch between the theoretical guarantees of confidential computing and the practical reality of how it is implemented.

This mismatch becomes especially relevant since many solutions in the domain of privacy-preserving data processing utilise confidential computing and its assumed security guarantees to achieve the envisioned privacy goals.
A few examples are: Mo et al.~\cite{CCforML} survey research publications that envision enhancing privacy and security for machine learning using confidential computing, Segarra et al.~\cite{CCforMedTec} propose to use confidential computing to protect the processing of medical data  and Khan et al.~\cite{CCforSensing} propose to use confidential computing to protect sensor data fusion.

%
However, while the former are primarily research efforts that build on the conceptually achievable trust assumptions, there is also the German certification body gematik, which mandates the use of confidential computing due to its strong security and privacy guarantees.
This is particularly vital because gematik requires the use of \acp{TEE} in a number of use cases, for example in the context of Confidential Healthcare Cloud (HCC)~\cite{gem3}, electronic patient records (ePA)~\cite{gem2} and sectoral identity provider (Sek-IDP)~\cite{gem1}. Infrastructure providers in these cases must be technically prohibited from creating user profiles, as well as from accessing underlying cryptographic material. 
Also referred to as provider exclusion.

Due to these stark discrepancies regarding trust assumptions between proposed solutions in research, commercial cloud offerings, and public certification bodies, this SoK paper aims to provide a comprehensive study of public cloud confidential computing services, with a particular focus on remote attestation mechanisms. 
Thereby, we analyse how major cloud providers structure their attestation flows, where trust is placed, and how these models compare to the original security principles of confidential computing technologies. 
Our systematisation of knowledge makes the following contributions:
\begin{itemize}
    \item development of a taxonomy for the evaluation of remote attestation frameworks in public cloud offerings;
    \item analysis and evaluation of the major public cloud offerings;
    \item examination of what confidential computing offerings are currently able to deliver and where they fail.
    \item Finally, we discuss possible root causes of the discrepancy between trust assumptions in research systems and the analysed commercial cloud offerings.
\end{itemize}

The rest of the paper is structured as follows:
In Section~\ref{sec:bac}, we provide the essential foundations for the SoK, including the threat model for confidential computing and some background on current hardware extensions (i.e., AMD \ac{SNP} and Intel \ac{TDX}) to enable cloud-based confidential computing.
This is followed by a discussion of related studies in Section~\ref{sec:rw}.
Section~\ref{sec:tax} presents a taxonomy for analysing the confidential computing cloud offerings under consideration.
Next, Section~\ref{sec:ana} details the actual analysis of the major public cloud offerings, while Section~\ref{sec:con} draws final conclusions.  

\section{Background}
\label{sec:bac}

In this section, we start by describing the threat model of confidential computing, based on which we highlight the shortcomings of the analysed public cloud offerings in Section~\ref{sec:ana}.
Furthermore, we introduce the \ac{RATS} architecture to formally define the main actors in the attestation process and, to make the analysis more comprehensive. Finally, we provide more technical details on the underlying technologies, namely \acl{TPM}, AMD~\ac{SNP}, and Intel~\ac{TDX}.

\subsection{Threat Model}
\label{sec:threatmodel}
As a basis for this study, we assume a typical threat model for \acp{TEE}.
We consider the cloud provider to be untrusted, as proposed by various research systems~\cite{baumann15haven,arnautov16scone,tsai17graphene-sgx}. 
Furthermore, an attacker may have complete control over the software environment, including all privileged system software (i.e., hypervisor, host operating system), with the goal of breaking the confidentiality or integrity of the code running in the \ac{TEE}.

Availability threats, such as interfering with \ac{TEE} execution, are not of interest: By design, the cloud provider and the surrounding system software are able to stop execution of \acp{TEE} at any time.
Besides that, we consider side-channel attacks to be out of scope and assume that enclaves do not have any security-related vulnerabilities that could lead to data leakage or integrity violations. 
We also fully trust the design and correct implementation of all the hardware and software mechanisms necessary to implement the \ac{TEE}, including all necessary cryptographic operations.

\subsection{Remote Attestation}

\begin{figure}[tb]
    \centering
    \includegraphics[width=0.8\columnwidth]{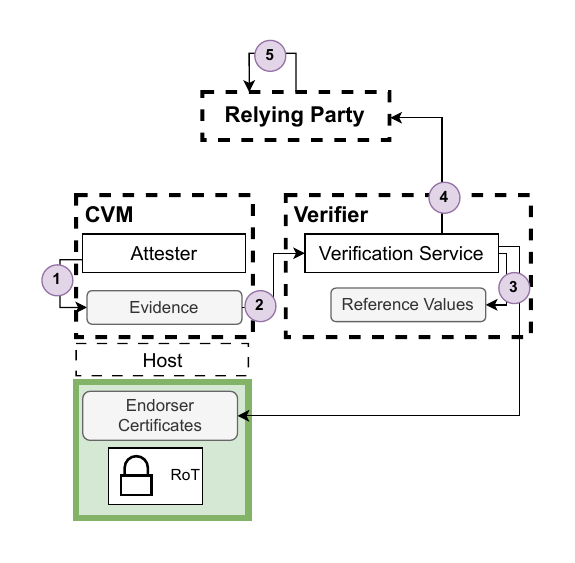}
    \caption{RATs Architecture workflow: (1) Attester produces evidence; (2) evidence is send to Verifier; (3) Verifier validates evidence based on Reference Values and Endorsements; (4) Attestation Result is shared with Relying Party; (5) Relying Party makes trust decision.}
    \label{fig:rats}
\end{figure}

Remote attestation, as defined by the \ac{RATS} architecture~\cite{RFC9334}, is a process in which one entity (Attester) produces believable information (Evidence) about itself to enable a remote entity (Relying Party) to decide whether to consider the Attester trustworthy. 
Figure~\ref{fig:rats} depicts this architecture.
The process is facilitated by an additional party, called Verifier, who appraises the Evidence and creates Attestation Results. 
The Relying Party uses these results later on to make trust decisions. 
Additional roles include Endorser and Reference Value Provider, who supplies Endorsements and Reference Values, respectively, to the Verifier. 

The \ac{RATS} architecture~\cite{RATSOverview} involves a two-stage appraisal procedure, verifying the Evidence and the Attestation Results. 
For the first step, the Verifier applies policies and supply chain input, such as Endorsements and Reference Values, to create Attestation Results from the Evidence provided by the Attester. 
During the second part of the procedure, the Relying Party applies its own policies to the Attestation Results associated with an Attester's Evidence, which originates from a trusted Verifier. 
This appraisal leads to trust decisions regarding the Attester. 

\subsection{Trusted Platform Module}
A \acf{TPM}~\cite{TPM2Spec} is a secure crypto-processor designed to provide hardware-based cryptographic functions and tamper-resistant storage. 
It can be used for random number generation, secure generation and storage of cryptographic keys, data encryption (symmetric and asymmetric), as well as cryptographic hash functions. 
Its main use case is typically measured boot, where the \ac{TPM} acts as \ac{HRoT}, hashing the boot components of a system sequentially, and storing these hashes in immutable chains in \ac{TPM}'s \acp{PCR}. 
\acp{PCR} cannot be directly written to; instead, values are ``extended'' through a hashing process, creating a cumulative record of the system's state.
The \ac{TPM} acts as a Root of Trust for reporting as well, providing integrity data that can be queried by external verifiers, boot loaders, kernels, or software agents. 
Secure Boot~\cite{juniper2020securemeasuredboot} on the other hand, provides active enforcement against untrusted binaries by verifying their digital signatures and allowing only signed code to execute during startup. 
Furthermore, a \ac{TPM} provides secure key generation, storage, and management, ensuring that cryptographic keys remain protected from software-based attacks. 
Keys handled by a \ac{TPM} are typically non-exportable, meaning private keys never leave the \ac{TPM} and can also be made persistent in the \ac{TPM}'s \ac{NVRAM}. 
In a physical \ac{TPM}, NVRAM is a physical component integrated into the \ac{TPM} chip itself and provides tamper-resistant storage, ensuring that sensitive data remains protected even when the system is off.

In case the host system is used to deploy multiple \acp{VM} that cannot share the same physical \ac{TPM} for security reasons, virtual \acp{TPM} (vTPMs)~\cite{Berger2006vTPMVT} can be used instead. 
These are software-based emulations of the physical \acp{TPM}, tailored for virtualised settings, allowing VMs to access trusted computing features without dedicated hardware \acp{TPM}.
vTPMs can replicate the full range of \ac{TPM} functionalities, but they come with certain trade-offs. 
Their security guarantees are generally less robust than those of physical \acp{TPM}, and they have a larger attack surface. 
This increased vulnerability stems from the necessary involvement of the hypervisor in managing the vTPMs, which introduces additional complexity and potential points of compromise.
An example of such compromise is the simulated \ac{NVRAM}. 
Unlike hardware \acp{TPM}, where \ac{NVRAM} is physically secured within the chip, vTPMs typically simulate it as a file stored on the host system. 
While this file is often encrypted to protect confidentiality, it does not inherently provide integrity protection against tampering or rollback attacks, so the stored key handles, which can eventually be used for attestation purposes, are susceptible to leakage or corruption.

\subsection{Intel TDX}
\acf{TDX}~\cite{tdxdemystified} is Intel's latest confidential computing technology, designed to provide robust isolation at the VM level, by deploying \acp{TD}, which are protected from the host operating system, hypervisor, and other VMs running on the same hardware.
\acp{TD} are hardware-isolated VMs whose memory remain encrypted with a unique key during runtime leveraging Intel's \ac{TME-MK}.

Verifying the integrity and authenticity of a \ac{TD} is possible by performing remote attestation over the \ac{TD} Quote, which is \ac{TDX}-signed proof about the state of the \ac{TD} and the underlying platform.
The \ac{TDX} module~\cite{IntelTDXGuide} records measurements of the \ac{TD}'s state before and after launch; before its initialisation measurements pertaining to the guest's initial configuration and firmware image are recorded on the \ac{MRTD}, while after its initialisation the four \acp{RTMR} can be used and extended by the virtual firmware (e.g., TDVF or TD-Shim)~\cite{IntelTDX2023} to record \ac{TD}'s runtime state. 
The first register typically records the firmware data, i.e. TD-Shim configuration, the second one includes the measurements of OS code (e.g., kernel image), the third one reflects the boot configuration (e.g., command line parameters, initrd) and the last one can be leveraged by runtime integrity monitoring systems to record events triggered by userland applications and the system after boot. 

The \ac{TD} Quote~\cite{IntelTDXOverview, IntelTDXDocumentation} is essentially a signed \ac{TD} report, hence a report must be requested first from the \ac{TDX} module. 
This contains guest-specific information, platform security version, a \ac{MAC} for integrity protection, as well a 64-byte user-provided \textit{report\_data}, which typically is a nonce provided by the attestation service. 
After the report has been created, it is sent to the \ac{TDQE}, an \ac{SGX} enclave.
The \ac{TDQE} verifies the report locally by checking the \ac{MAC} and then signs it with its \ac{AK} to create the \ac{TD} Quote.

Besides the signed report, the TD Quote also contains the Attestation Key Certificate (signed by the Provisioning Certification Key) and the \ac{PCK} Certificate (signed by Intel). 
Similar to the \ac{DCAP} attestation flow~\cite{IntelSGXDCAP2023} the generation of the Attestation Key occurs during the provisioning of the \ac{TDQE}, where after the creation of this key pair, the \ac{PCE} signs the public part of the \ac{AK} using the \ac{PCK}.
An Intel-provided \ac{PCK} Certificate, containing the public key corresponding to the \ac{PCK} signature, is made available to the \ac{QE}.
The \ac{PCK} Certificate's trust is rooted in an Intel \ac{CA}.

\subsection{AMD SEV-SNP}
AMD \acl{SNP}~\cite{AMDSEVSNP} is the third iteration of AMD's \ac{SEV} technology introducing memory integrity protection to the VM-based \acp{TEE}. 
AMD \ac{SEV} employs a unique encryption key for each VM to protect its memory contents during runtime, leveraging the AMD Secure Processor (AMD-SP), a dedicated on-chip component that operates independently of the hypervisor.
Within the SEV-SNP architecture, the \ac{TCB} is comprised by the hardware and several firmware components, such as the AMD-SP API and the CPU microcode patch, which are considered upgradeable. 
AMD SEV-SNP also enables the VM owner to further partition its address space into four hierarchical levels, providing fine-grained security control between different components of the software stack through hardware-isolated abstraction layers within the VM. 
These are called \acp{VMPL}~\cite{TUMSEVAnalysis, KollendaSEVTDX, EnclaiveVMPL} and are numbered \ac{VMPL}0 to \ac{VMPL}3, with \ac{VMPL}0 being the most privileged and \ac{VMPL}3 the least.
Higher privileged \acp{VMPL} can implement secure services~\cite{AMDSVSM,LinuxSVSM} for lesser privileged \acp{VMPL}. 
For example, \ac{VMPL}0 can be used to run security-critical code or implement a vTPM, isolating it from potentially compromised lower-privilege levels.
\acp{VMPL} can also enable a guest to contain its own management layer running at a high \ac{VMPL}, controlling the permissions on its other pages. 
This allows secure virtualisation of a security-enforcing hypervisor within the guest VM itself, thereby enabling secure nested virtualisation~\cite{Hecate}.

To verify the integrity and authenticity of an SNP-protected VM, a remote party can request an attestation report that will eventually be generated by the AMD-SP.
The report contains several key components~\cite{AMDSEVSNPABI}; namely the \ac{TCB} version which includes information about the firmware and hardware versions, the VM's initial state cryptographic measurements which reflect VM's initial memory contents, configuration and 64-bytes of user supplied data, such as a nonce to prevent replay attacks.
The report can be signed by one of the following attestation keys, which are both dependent on the \ac{TCB} version;
the \ac{VCEK}, which is derived from chip-unique secrets making it platform specific, or the \ac{VLEK}, which is derived from a seed maintained by the AMD \ac{KDS}. 
Each \ac{CSP} that enrolls with AMD has dedicated \ac{VLEK} seeds, allowing for workload movement within the \ac{CSP} realm.
Both \ac{VLEK} and \ac{VCEK} are signed by \ac{ASK}, that in turn is signed by \ac{ARK}, which is the master key behind the AMD certificate authority. 
The latter is self-signed and kept private by AMD.
The attestation report also contains other data, like the \textit{CHIP\_ID}, a unique identifier for the AMD processor chip (used with \ac{VCEK}) and the \textit{CSP\_ID}, the \ac{CSP} identifier (used with \ac{VLEK}). 
It can also hold the digest of an ID key, which can be passed as an argument during the launch of the VM to uniquely identify it, as well as the \ac{VMPL} identifier to report on which privilege level the application that requested the report runs on. 

Regarding the VM's initial state measurement, which is also included in the attestation report, it typically includes only the hash of the virtual firmware because it corresponds to the first virtual firmware volume loaded during the boot process. 
This limited measurement is insufficient for ensuring the integrity of the entire VM, as it does not account for the operating system or root filesystem state~\cite{revelio,DecentriqSEVSNP}. 
To address this limitation, measured direct boot has been implemented as a series of patches to the hypervisor, QEMU, and the virtual firmware (i.e., OVMF). 
This approach creates space in the firmware binary to store a table containing hashes of the kernel, initial RAM disk (initrd), and kernel command line. 
When QEMU boots a VM, it calculates hashes for these components and injects them into the special table. 
OVMF then measures each component during startup and compares the measurements with those in the designated table. 
This ensures more comprehensive representation of the VM's initial state, as their cryptographic hashes are reflected in the launch measurement in the attestation report.

\section{Related Work}
\label{sec:rw}

A wide range of surveys exist on \acp{TEE}.
Sabt et al.~\cite{sabt2015trusted} describe the key properties and concepts of \acp{TEE} to enable a clear definition of it.
Munoz et al.~\cite{munoz2023survey} analyse vulnerabilities in \acp{TEE}, whereby they enlist a wide range of possible software-based, side-channel and architectural attacks.
Li et al.~\cite{li2024sok} systematically examine the design choices of \acp{TEE} and discuss the security implications.

With regard to attestation,
Ménétry et al.~\cite{menetrey2022attestation} compare attestation mechanisms of different \ac{TEE} implementations.
Their work provides an overview of attestation mechanisms, which is an important basis for our analysis of commercial offerings. 
Niemi et al.~\cite{niemi2022towards} provide a survey of attestation frameworks.
Their work classifies open-source enclave projects, while we study \ac{CVM}-based offerings. 
Similarly to our work, Scopelliti et al.~\cite{scopelliti2024understanding} analyse trust relationships in confidential computing offerings in the cloud.
They define different attestation levels and apply them to the AMD \ac{SNP} based \ac{CVM} offerings by \ac{AWS}, Azure and \ac{GCP}.
Their analysis covers a subgroup of the offerings we study.
Furthermore, our work provides a more fine-grained analysis of the underlying architecture and the implications for remote attestation. 

\section{Taxonomy}
\label{sec:tax}

Our taxonomy provides the means to investigate whether confidential computing offerings truly allow to keep the infrastructure provider outside of the trust model. 
Prior to this work, Niemi et al.~\cite{niemi2022towards} defined eight questions for the evaluation of attestation frameworks. 
While they take the theoretical foundations of attestation into account, we use the current state-of-the-art as the baseline for our study.
Besides that, Scopelliti et al.~\cite{scopelliti2024understanding} provide a classification of \ac{CVM} offerings in different attestation levels.
While this work provides valuable criteria for analysing attestation, our taxonomy gives a broader analysis of aspects contributing to hardware-based confidential computing, such as different RoT participating in the attestation process and the establishment of a unique target identity.
We aim to apply our taxonomy to two types of \ac{VM}-based execution contexts; containers and \acp{VM} on Intel \ac{TDX} and AMD \ac{SNP} platforms.
We consider those to be the target of the attestation.
Our taxonomy consists of three categories: \textit{Identity Establishment}, \textit{Attestation-collateral Metrics} and \textit{\ac{TCB}}.

\subsection{Identity Establishment}

The owner of a confidential computing secured execution context expects to be able to bind a unique identity to the target, mitigating the risk of impersonation attacks. 
Furthermore, there must be at least one \ac{HRoT}, which is involved in collecting, reporting and signing the evidence. 
To verify the integrity and authenticity of such evidence, the verifier will need to validate the signature of the \ac{HRoT}. 
Additionally, there may be further Roots of Trust, acting as trust anchors for collecting and reporting runtime measurements.
In the following sections, we use RoT as the general term representing software-based components, such as vTPMs, while HRoT solely refers to hardware-backed entities and their respective keys, like in the case of \ac{TEE}s.

To summarise, we aim at answering the following questions:
\begin{enumerate}
    \item Which hardware identity is involved in collecting and signing the evidence?
    \item How is the identity of the target established?
    \item How is the target identity linked to the evidence?
\end{enumerate}

\subsection{Attestation-collateral Metrics}

Attestation-collateral metrics (in the following sections called metrics) are used for defining the target's properties, so that decisions on integrity and authenticity are possible~\cite{niemi2022towards}. 
These metrics can contain one or more of the following: cryptographic hashes of the target's context (e.g., files, data and events), identities linked to the target, allowing for \textit{hash-based} or \textit{key-based attestation}, respectively, or configuration data of the target. 
The hash-based metrics can be collected during loading time of the target or during its runtime.
Hash-based metrics are static, meaning that they are only collected by the RoT during the initialisation of the target's context and are immutable after that, while key-based metrics are dynamic.
They can be extended based on user-provided policies and can be event-triggered if paired with a runtime-integrity monitoring framework (i.e. \ac{IMA}).
The key-based metrics are set during the target configuration and reported by the RoT in the attestation collateral.

During the target configuration, some security policies can be enforced, i.e. secure boot, or usage of vTPM.
The enforcement of this can also be reported.

During the reporting of the attestation-collateral metrics, the verifier has the possibility to pass arbitrary data that will be cryptographically linked to the attestation report by the hardware in the attestation report.
The attestation report will contain the collected metrics, the verifier-provided data and possibly information about the \ac{TEE} platform.
It will be signed by the \ac{HRoT}.

Therefore, the following properties will be evaluated:
\begin{enumerate}
    \item What are the load time metrics?
    \item If runtime metrics are available, which entity is responsible for collecting them?
    \item Is it possible to link user-provided data in the attestation collateral during reporting?
    \item Is the configuration of security policies supported and is it part of the attestation collateral?
    \item If key-based attestation is used: How is it enforced, and which are the RoTs for this?
    \item Are the collected metrics attestable and reproducible?
\end{enumerate}

\subsection{Trusted Computing Base}

According to the threat model, which was described in Section~\ref{sec:threatmodel}, the \ac{TCB} contains the \ac{TEE} itself and the hardware it runs on.
Furthermore, we use the term trusted stakeholders to describe entities that need to be trusted during attestation.

We differentiate between components that need to be implicitly trusted, and components, that can be verified. 
Implicit trust is typically at least required in the \ac{TEE}'s hardware and firmware. 
The guest expects that all other components within the \ac{TCB} are measurable and attestable. 

Therefore, the following questions will be investigated:
\begin{enumerate}
    \item Which components can be verified by a relying party and which have to be trusted implicitly?
    \item What stakeholders are involved in the attestation?
    \item Is it possible to configure the \ac{TCB}?
\end{enumerate}

\section{Analysis}
\label{sec:ana}

\begin{figure}[tb]
    \centering
    \includegraphics[width=0.8\columnwidth]{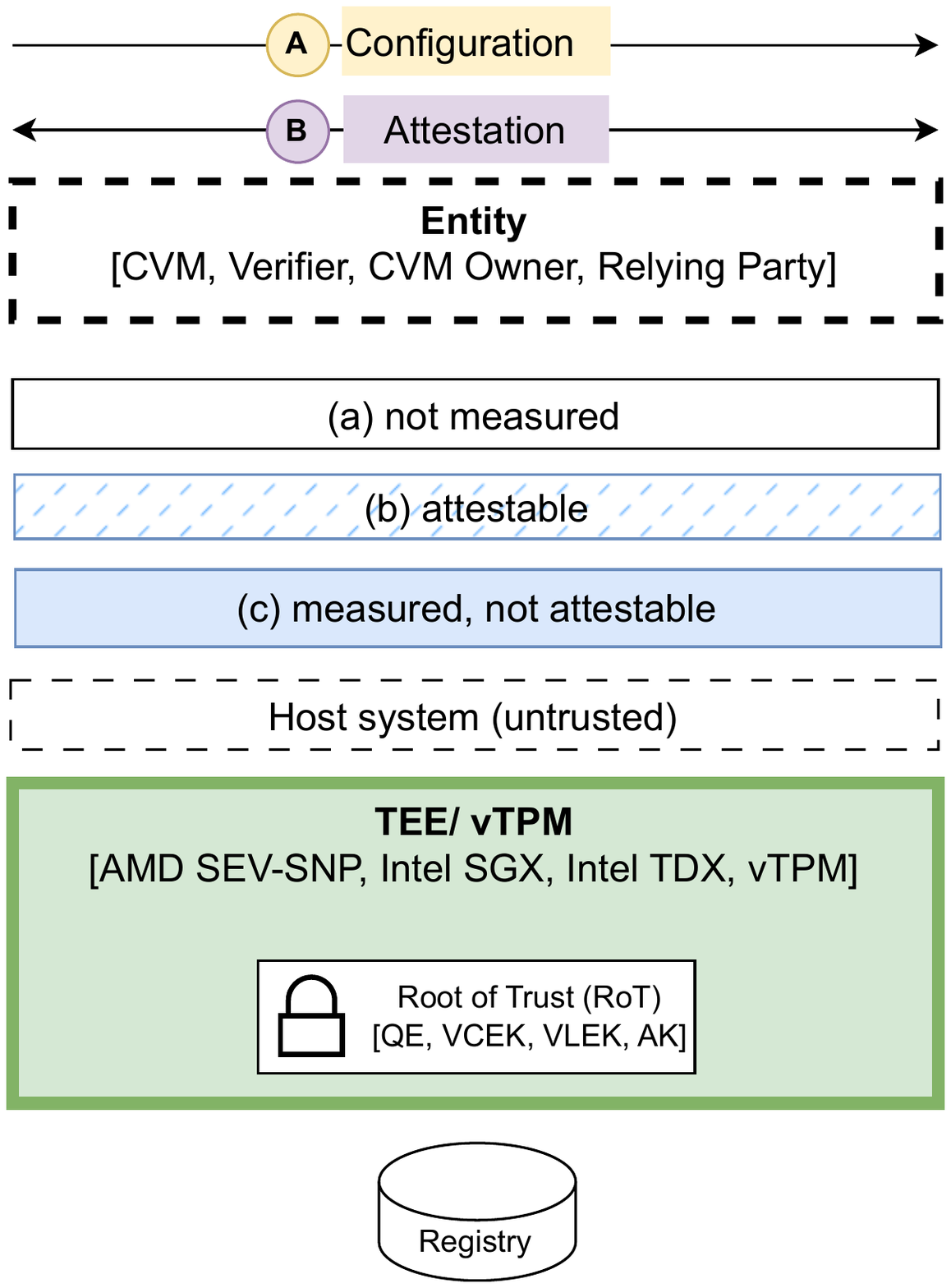}
    \caption{Graphical notation for describing architectures used throughout the paper. We differentiate between (a) \ac{TEE} components, which are not measured by a RoT, (b) components which are attestable, as they are measured and the measurement can be reproduced, and (c) components, which are measured, but due to closed source software not attestable, as measurements cannot be reproduced.}
    \label{fig:legend}
\end{figure}

Based on the described taxonomy, we evaluate different \ac{CVM} and confidential container offerings in the domain of public clouds.
Figure~\ref{fig:legend} depicts the graphical notation we use in the following.

We target \ac{IaaS} offerings, focussing on hardware-based \ac{CVM} offerings, also analysing confidential container solutions enabled by \acp{CVM}.
Based on market share, we selected Azure (20\%), AWS (31\%) and Google Cloud (12\%) as the most relevant \acp{CSP}~\cite{SynergyResearchGroup2025}. 

After analysing the different solutions individually, we provide a comparison of the different offerings.
Finally, we discuss possible root causes that may be responsible for the identified deficiencies.

\subsection{Microsoft Azure}
Microsoft Azure has the following \ac{CVM} offerings:
\begin{enumerate}
    \item \acp{CVM} using AMD \ac{SNP}
    \item Confidential containers running in AMD SEV-SNP CVMs
\end{enumerate}

Intel \ac{TDX} based \acp{CVM} are currently only available as a preview version. 
Therefore, this offering is not being considered.

\subsubsection{AMD \ac{SNP} \acp{CVM}}

Azure \acp{CVM} do not have direct access to the hardware, instead, they rely on a paravisor called \ac{HCL}.
Azure is developing an open source version of it, called OpenHCL~\cite{PulapakaOpenHCL}, but at the time of writing it is not being used in their offerings.
When attesting a \ac{CVM}, three different components must be considered:

\begin{enumerate}
    \item the \ac{CVM} itself;
    \item the vTPM, part of the \ac{CVM};
    \item \ac{MAA} Service, the verifier provided by Azure.
\end{enumerate}

\begin{figure}[tb]
    \centering
    \includegraphics[width=\columnwidth]{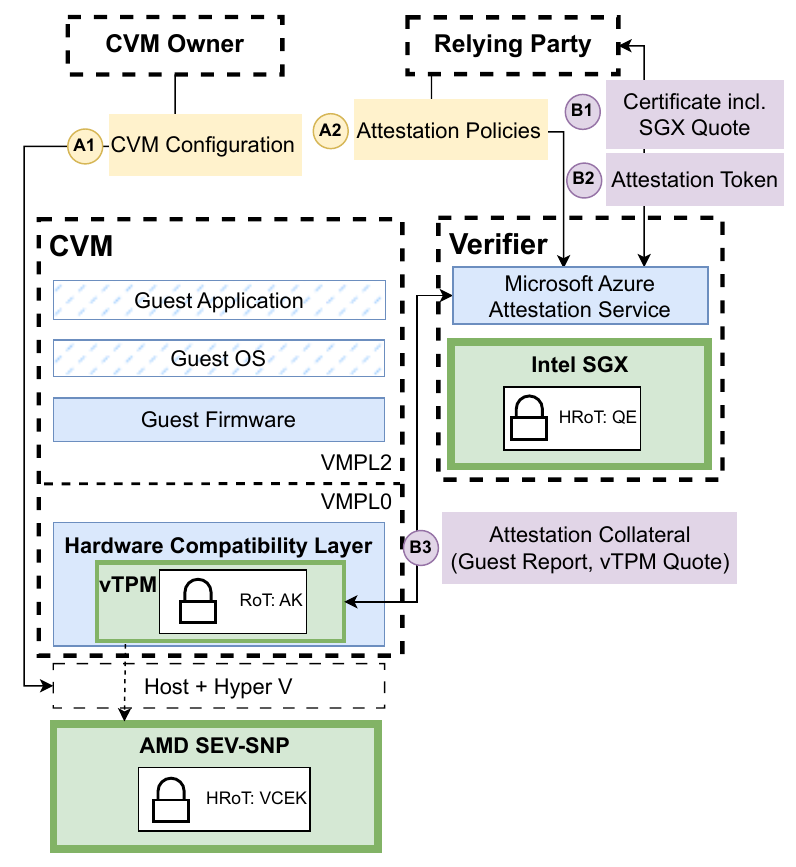}
    \caption{Azure \ac{SEV} \ac{CVM}: Architecture. \ac{MAA} is considered as the verifier but is optional and could be replaced.}
    \label{fig:azure-snp-cvm}
\end{figure}

\begin{figure}[tb]
    \centering
    \includegraphics[width=\columnwidth]{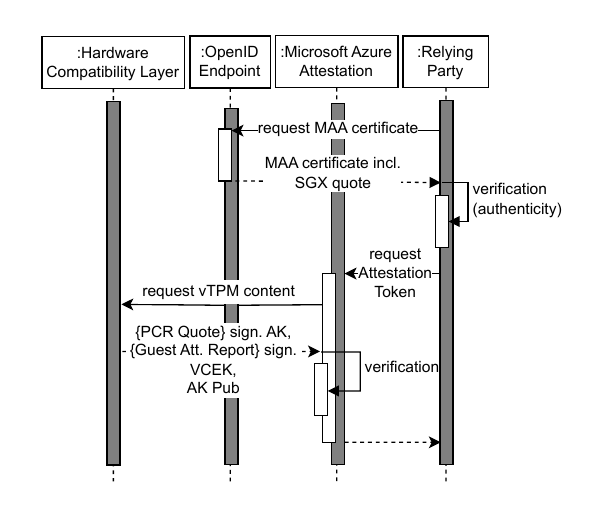}
    \caption{Azure \ac{SEV} \ac{CVM}: Sequence diagram of attestation when using \ac{MAA}.}
    \label{fig:azure-cvm-sq}
\end{figure}

Figure~\ref{fig:azure-snp-cvm} shows the architecture of the \ac{SNP} \ac{CVM} offering provided by Azure.
It considers the usage of \ac{MAA} as verifier, which is optional.
The retrieval of the attestation collateral (B2) by an arbitrary verifier remains the same; it is requested from the vTPM hosted inside the \ac{CVM}, where the guest attestation report is cached.
In the setup phase, the owner is able to configure the target with regard to the used base image and security policies, which allow to enable vTPM and secure boot, respectively~\cite{MicrosoftArm}. 
Further detailed configuration options are not provided for secure boot.
Furthermore, the relying party can configure the attestation policies of the \ac{MAA} service, which is acting as verifier and running in an \ac{SGX} enclave.
The configuration is done by providing attestation policies, which are consulted by \ac{MAA} when verifying a report. 

\acp{PCR} of the vTPM can be extended for runtime attestation, if enabled.
The vTPM follows the \ac{TPM} 2.0 specification~\cite{TPM2Spec}.
Azure uses the term `runtime claims'~\cite{AzurevTPM} to summarise Azure-specific \ac{NVRAM} content, which covers the \ac{CVM}'s security configuration and the vTPM's \ac{AK}.
A hash of this content is linked to the attestation report.

A guest attestation report is requested during the loading time by the \ac{HCL} and cached in the vTPM.
The vTPM furthermore allows to retrieve signed \ac{TPM} Quotes from the \ac{NVRAM}.
Further content of the \ac{NVRAM}, such as the \ac{AK} public and \ac{AK} certificate, can be read as well.
The \ac{AK} is used for signing the quotes, verification of the quote's authenticity can therefore be achieved using the \ac{AK} certificate.  
Verification of the guest attestation report's authenticity can be done after retrieving the \ac{VCEK} and \ac{AK} certificates from the Azure Instance Metadata Service and the certificate chain from AMD's \ac{KDS}~\cite{azurevTPMtutorial}.

The \ac{CVM} can be attested by consulting the Azure provided verifier \ac{MAA} as shown in Figure~\ref{fig:azure-cvm-sq}.
A self-signed certificate of the \ac{MAA} can be retrieved from a provided \ac{JWKS} endpoint.
The certificate is extended with the \ac{DCAP} attestation quote of the MAA service, allowing to attest the authenticity of the service.
On request of the relying party, the \ac{MAA} retrieves the guest attestation report from the \ac{CVM}'s vTPM.
The response is a so-called attestation token, the result of evaluating the previously defined attestation policies.
The attestation token is signed with \ac{MAA}'s private key.

\paragraph{Identity Establishment}

The hardware identity of the \ac{SP}, \ac{VCEK}, is responsible for collecting and signing the evidence.
The owner is not able to pass a key-based identity during boot so uniquely identifying the \ac{CVM} in that way is not possible. 
Its hardware-recorded measurement cannot serve as a unique identity either for two main reasons; the measured envelope is not fully representative of the \ac{CVM} state and many of its components are closed source and hence cannot be reproduced, as we will explain in the next section.

The vTPM~\cite{AzurevTPM} uses a persistent \ac{AK}, with a certificate issued by Azure, for signing evidence.
The \ac{AK} is linked to the \ac{CVM} by including the hash of the `runtime claims' in the attestation report.
Although we were able to verify this certificate against some key provided in an Azure FAQ, the corresponding \ac{CA} is not well documented.
It is not transparent to the user how the \ac{AK} is generated, so it is unclear if it is based on a random process.
Though an \ac{EK} exists, we were unable to confirm a linking of the \ac{AK} to it.

The \ac{MAA} service, acting as a verifier, retrieves the attestation report from the vTPM and provides a signed attestation token to the relying party.
It runs in an \ac{SGX} enclave and has its hardware identity rooted in the DCAP \ac{QE}.
It is already alive during the setup of the \ac{CVM}.

Depending on the chosen attestation path, different identities are involved in reporting.
The measurements are initially collected by the \ac{SP} and signed with a \ac{VCEK}.
During runtime, metrics are collected by the vTPM running in the \ac{HCL}, which uses a persistent key, the \ac{AK}, for signing.
The identity of the vTPM is linked to the \ac{CVM}'s \ac{VCEK} by including the \ac{AK}'s public part in the attestation report.
If verification is conducted by using \ac{MAA}, this service is responsible for signing the attestation token and verifying the identities used to record the attestation collateral. 

\paragraph{Attestation-collateral Metrics}

The user can configure user security policies, which are limited to enabling secure boot and vTPM.
These can be reported to the relying party by the \ac{MAA} service or retrieved from the vTPM's \ac{NVRAM}.
This content cannot be authenticated, as it is unsigned; only the quotes are signed by the \ac{AK}. 

During loading time, the initial firmware is measured.
The measurement cannot be attested, as the \ac{HCL} and guest firmware are closed source and therefore not reproducible.
If the vTPM is used, runtime measurements can be collected by extending the \ac{PCR} registers.
The identity of the vTPM, the \ac{AK}, is linked to the attestation report via including a hash in the \texttt{report\_data} field.
The runtime measurements are in principle attestable, but require implicit trust in the \ac{HCL}, where the vTPM is emulated.

The \ac{MAA} service, acting as verifier, allows custom data to be included in an attestation token for ensuring freshness of the token. 
This data solely allows to customise the attestation token and ensure its freshness, but not of the attestation report. 
The service runs in an \ac{SGX} enclave and is attestable with regard to authenticity.
The certificate of \ac{MAA} includes the \ac{SGX} attestation quote of the enclave, which contains the hash-based \texttt{MRENCLAVE} value of the \ac{MAA}.
The measurement cannot be reproduced by the \ac{MAA} users, as the implementation is closed source.
Validating the integrity of the measurement is only possible by comparing it to an Azure-provided value.

\paragraph{\ac{TCB}}

The \ac{TCB} of the \ac{CVM} is composed of the software running on the \ac{SP}, \ac{HCL}, guest firmware, and guest OS.
Since guest firmware and \ac{HCL} are closed source and the measurements are therefore not reproducible, the user is not able to attest if the guest has been booted correctly.
Implicit trust in these components is therefore necessary.
The guest OS can be measured by the vTPM and attested, if the \ac{HCL} is trusted.

If the user delegates verification to \ac{MAA}, Azure participates in the attestation and has to be trusted implicitly, as the service's integrity can only be attested using an Azure-provided reference value. 
Reproducing the measurement is not possible, as the service is build from a closed-source implementation.
Configuration of the \ac{TCB} is limited to selecting the target guest OS image, the firmware cannot be customised.

\paragraph{Conclusion}

The \ac{SP} \ac{CVM} solution offered by Azure is able to provide the following:
\begin{itemize}
    \item Retrieval of loading-time metrics, either by reading a cached guest attestation report from a vTPM or alternatively by requesting Attestation Tokens from a verifier called \ac{MAA} provided by Azure.
    \item A vTPM inside of the \ac{CVM} can be utilised for collecting runtime claims. The \ac{AK} of the vTPM, used for collecting and signing, is bound to the \ac{CVM} via a hash in the \texttt{report\_data} field of the attestation report.
\end{itemize}

The following shortcomings do exist:
\begin{itemize}
    \item The \ac{CVM} has no direct access to hardware, so that attestation reports cannot be directly retrieved. Instead, it relies on Azure's \ac{HCL} as paravisor.
    \item The owner is unable to uniquely identify the target, as no identity data can be passed and the measurements do not allow unique identification of the target either. 
    \item Interactions with the \ac{MAA} require implicit trust in Azure.
    \item The vTPM is solely implemented in software, using Azure closed-source code. The user is unable to attest this component.
    Therefore, Azure needs to be implicitly trusted.
    \item Freshness of attestation reports is not given, as they are requested by the \ac{HCL} at loading time and cached in the vTPM. No custom data can be provided when retrieving the guest attestation report from vTPM.
    \item The \ac{TCB} cannot be attested, as the \ac{HCL} and guest firmware are closed source and measurements are therefore not reproducible.
\end{itemize}

\subsubsection{Confidential Containers}

\begin{figure}[tb]
    \centering
    \includegraphics[width=\columnwidth]{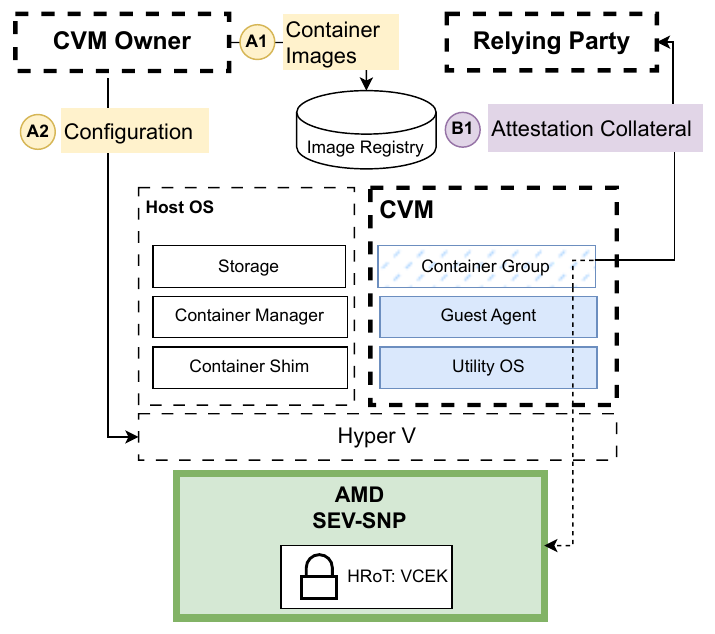}
    \caption{Azure \ac{SNP} \ac{CVM} Container Group architecture.}
    \label{fig:azure-container}
\end{figure}

Azure offers confidential container groups running in \ac{SNP} \acp{CVM}~\cite{azureConfContainers} based on Parma~\cite{johnson2024confidential}.
Parma provides the possibility to launch a group of containers within a \ac{CVM} with verifiable execution policies that are enforced by the guest agent.
The execution policies link the image layers via \textit{dm-verity} hashes~\cite{dmverity}, so that tampering can be detected by the guest agent prior to launch.
Furthermore, the execution policies define commands, entry points, and environment variables that are enforced by the guest agent.
The \ac{SP} measures the \ac{UVM}, composed of guest OS and guest agent, at loading time.
The policies are included in the \texttt{host\_data} attribute of the guest attestation report, allowing the verifier to validate if the container group has been launched with the expected configuration.
The container shim, running on the host, passes the policy to the guest agent, which measures the policy and compares it to the value encoded in the attestation report.

Azure adopts Parma for their confidential container group offering, shown in Figure~\ref{fig:azure-container}.
Similar to the Parma model, the owner is able to configure execution policies, which are to be enforced by the guest agent and linked to the guest attestation report.
However, configuration of the \ac{UVM}, such as OS or firmware, is not possible.
Attestation reports can be requested by privileged containers directly from the hardware.

\paragraph{Identity Establishment}

The \ac{SP} is responsible for collecting the evidence, attestation reports are signed with a \ac{VCEK}. 
The identity of the target cannot be established by providing an key-based identity before launch. 
As discussed in the next section, verifying the measured envelope is not feasible, making identification through measurements equally impossible. 

\paragraph{Attestation-collateral Metrics}

The launch measurement of the \ac{CVM} cannot be attested, as the \ac{UVM} is closed source.
A hash of the \ac{CCE} policies, which were configured prior to launch, are linked to the attestation report.
However, as the underlying \ac{UVM} is not attestable, it is unclear if the execution policies are actually enforced.
Attestation of the \ac{CCE} policies is therefore only possible if the underlying guest agent is implicitly trusted.

\paragraph{\ac{TCB}}

The \ac{TCB} of a Confidential Container Group is composed of the \ac{SP}, the \ac{UVM}, and the container image layers.
It is not possible to attest the \ac{UVM}, therefore implicit trust in this component is required.
As the \ac{UVM} is provided by Azure and cannot be verified, the \ac{CSP} needs to be trusted, which contradicts the trust model we described in Section~\ref{sec:threatmodel}.
Attestation of the \ac{CCE} policies, covering also container image layers, is possible, but only if the \ac{UVM} is trusted to enforce the policies.
The \ac{TCB} can only be configured with regard to the container image layers, selecting the underlying \ac{UVM} is not possible.

\paragraph{Conclusion}

Azure's confidential container groups have the following properties in regard to attestation:
\begin{itemize}
    \item Reports can be retrieved from hardware.
    \item \ac{CCE} policies allow to define the container image layers and behaviour of the container group in an attestable manner.
\end{itemize}

It has the following shortcomings:
\begin{itemize}
    \item No unique identity of the container group is established.
    \item The \ac{UVM} cannot be attested, as the implementation is closed source. Therefore, it has to be trusted to enforce the policy.
    \item Trust in the \ac{CSP} is required as the \ac{UVM} is provided by Azure.
    \item Configuration of the \ac{UVM} is not possible.
\end{itemize}

\subsection{\acf{AWS}}

\ac{AWS} offerings can leverage \ac{SNP} or the \ac{AWS} Nitro System for confidential computing. 
Nitro Enclaves~\cite{AmazonNitro} are managed and enforced by the Nitro Hypervisor, leading to Amazon both acting as hardware vendor and cloud provider.
As our trust model assumes that the infrastructure provider does not need to be trusted, we therefore only consider the \ac{SNP} offering.

\subsubsection{AMD SEV-SNP CVMs}

\begin{figure}[tb]
    \centering
    \includegraphics[width=0.9\columnwidth]{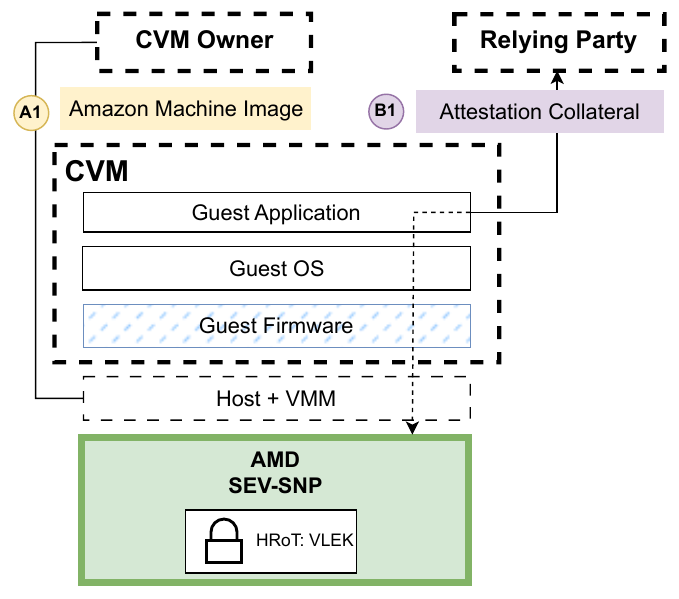}
    \caption{AWS \ac{SNP} \ac{CVM}: Architecture.}
    \label{fig:aws-snp-cvm}
\end{figure}

The workflow of setting up an AMD SEV-SNP \ac{CVM} is shown in Figure~\ref{fig:aws-snp-cvm}. As depicted, the owner configures the \ac{CVM} through an \ac{AMI}. The image is either derived from an existing \ac{AMI} or imported as a self-configured VM image. 
In both cases, the image is handled and modified by an \ac{AWS} closed source tool prior to launch and cannot be exported.
Providing an identity block or using custom firmware is not possible. 
After launch, attestation reports can be retrieved directly from the hardware. 
The verification is left to the user and can therefore be achieved by using open-source tools.

\paragraph{Identity Establishment}

The \ac{HRoT} is the \ac{SP}, the attestation reports are collected and signed with a \ac{VLEK}. 
The owner cannot bind identity data to the \ac{CVM} prior to launch for establishing the target's identity.
Since the components in the measured envelope do not cover the necessary context of the \ac{CVM}, uniquely identifying it by its measurement is not feasible either.

\paragraph{Attestation-collateral Metrics}

The measured envelope is solely composed of the virtual firmware.
Reproducing and verifying its measurement is possible, as the used firmware is open source~\cite{awsUefi}.
User-provided data can be linked on request of an attestation report, so that freshness can be ensured by the verifier.
Alternatively, the owner can bind a public key to the target. 
Runtime metrics are not available. 

The owner of a \ac{CVM} is able to configure the usage of a vTPM, more particular a NitroTPM~\cite{NitroTPM}, which runs outside of the \ac{CVM} and would require trust in the hypervisor.
As this is in conflict with our trust model, we do not consider this options.
Besides that, the configuration of secure boot is possible, which allows to customise the boot chain of the \ac{VM}.
However, we were not able to detect any changes in the attestation report's measurement, as only the virtual firmware, but not the kernel, is measured.
The UEFI variable store is loaded only after the \textit{sev init} is completed~\cite{awsuefipatch}. 
Therefore the configuration of secure boot does not influence the launch measurement.

\paragraph{Trusted Computing Base}

The used guest firmware is measured and open source~\cite{awsUefi}, hence the measurement can be reproduced and verified.
Attesting the guest OS is not possible due to two reasons.
Firstly, there exists no possibility to measure the guest OS, as the firmware cannot be customised and using the NitroTPM would require trust in the hypervisor.
Secondly, a measurement of the guest OS would not be reproducible, as the image cannot be retrieved by the customer after being modified by \ac{AWS} tooling.
As the owner is able to configure secure boot, configuring the \ac{TCB} is feasible to some extent, but this is not reflected on the launch measurement as we previously explained.
Modifications of the boot chain enforced by secure boot are therefore not captured by the measurement.

Due to the modified and non-attestable guest OS, \ac{AWS} becomes a trusted stakeholder on attestation.
This contradicts the standard trust model, which considers only the hardware vendor as trusted.

\paragraph{Conclusion}

We notice the following properties of the \ac{SNP} \ac{CVM} offering by \ac{AWS} in regard to attestability:
\begin{itemize}
    \item The guest OS is able to retrieve attestation reports using open source tools.
    \item Custom data can be linked to the target on request of an attestation report. 
    Thus, the verifier can ensure freshness and the owner can bind cryptographic material to the \ac{CVM}.
    \item The recorded measurement in the hardware attestation report can be verified, as the guest firmware is open source and can be reproduced.
\end{itemize}

However, we also noticed the following shortcomings:
\begin{itemize}
    \item The owner is not able to uniquely identify the \ac{CVM}. 
    This makes impersonation attacks possible, in which a tampered \ac{CVM} is used without notice by the user. 
    \item The usage of custom firmware is not possible.
    \item The relying party is not able to fully attest the TCB, as attesting the guest OS is not possible.
    \item Runtime measurements cannot be made without trust in the \ac{CSP}, as a NitroTPM would need to be used, which runs outside of the \ac{CVM} and is managed by the hypervisor.
\end{itemize}

\subsection{Google Cloud}

Google Cloud provides the following options for hardware-based confidential computing:

\begin{enumerate}
    \item CVMs using AMD SEV-SNP
    \item CVMs using Intel TDX
    \item Confidential containers, called `Confidential Space', which leverage AMD SEV CVMs
\end{enumerate}

As AMD SEV does not provide integrity protection over the \ac{CVM} or post boot attestation for its context, and because the solution solely relies on a cloud provider-managed vTPM as a RoT for recording/reporting, we only consider the offerings that are based on AMD \ac{SNP} and Intel \ac{TDX}.

\subsubsection{AMD SEV-SNP CVMs}\label{sec:gcp:amd}
\begin{figure}[tb]
    \centering
    \includegraphics[width=\columnwidth]{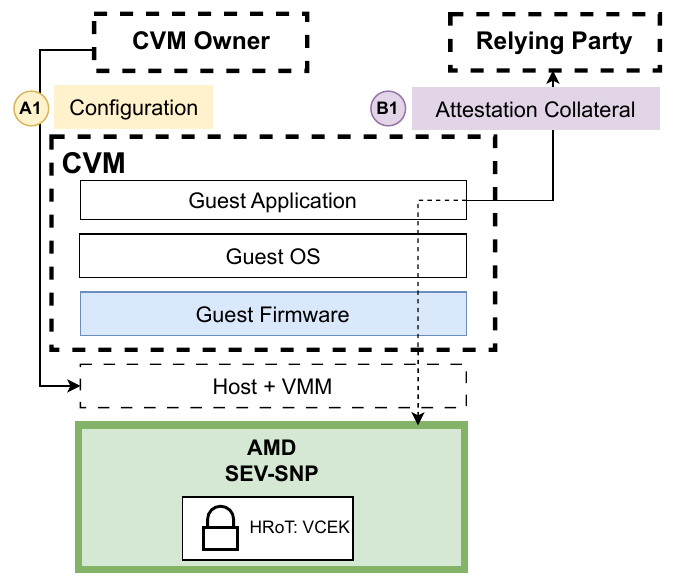}
    \caption{\ac{GCP} SEV-SNP \ac{CVM}: Architecture.}
    \label{fig:google-snp-cvm}
\end{figure}

The interactions on setting up and attesting a \ac{GCP} \ac{CVM} are shown in Figure~\ref{fig:google-snp-cvm}.
The owner is able to select a preconfigured, optionally customised, image~\cite{GoogleImage}.
The guest OS can retrieve the attestation reports directly from the hardware.
The verification of attestation reports can be achieved using open-source tools.

\paragraph{Identity Establishment}

The hardware identity to collect and sign the evidence is a \ac{VCEK}.
It is not possible for the owner to provide identity data prior to launch. 
Unique identification via the components in the measured envelope is not possible either, as guest OS and guest application are not covered in the measurement.

\paragraph{Attestation-collateral Metrics}

At loading time, the \ac{SP} and guest firmware are measured.
The collected measurement cannot be reproduced, as the firmware is closed source.
Verification is only possible by comparing the measurement with the reference values of Google~\cite{googleFirmware}.
It is not possible to conduct runtime measurements.
This would only be possible using the \ac{GCP} provided vTPM, but we were unable to find any information about its operation.
We therefore assume that it is running outside the \ac{CVM}, which requires trust in the \ac{CSP}.
As this contradicts the adopted trust model, we do not consider this as an option for collecting runtime metrics.
Guest attestation reports can be retrieved from the hardware using open-source tools, with the option to provide arbitrary data that is cryptographically linked to the attestation report.

\ac{GCP} allows to enrich a \ac{CVM} with further security policies, i.e. secure boot, vTPM, and integrity monitoring, whereby integrity monitoring reports the results of measured boot~\cite{googleShieldedVM}.
The enforcement is reported via a Google provided monitoring tool~\cite{googleMonitoring} and is not part of the attestation report, therefore it cannot be attested.
Secure boot can be activated or deactivated, respectively, and a behaviour in case of violation can be defined.

\paragraph{Trusted Computing Base}

The \ac{TCB} consists of the \ac{SP} firmware, guest firmware, and guest OS.
The guest firmware is owned by Google and closed source, measurements are therefore not reproducible.
The guest OS is not attestable without using the Google provided vTPM, which does not conform with our trust model.
Given that the guest firmware is closed source, implicit trust in Google is required since its implementation cannot be validated.
Customising the \ac{TCB} is limited to configuring the OS based on a Google-provided image.
The firmware cannot be customised.

\paragraph{Conclusion}

The \ac{SNP} \ac{CVM} offering provides the following assurances:
\begin{itemize}
    \item The guest OS can interact with the \ac{HRoT} directly. 
    \item Custom data can be provided on request of a report, allowing to bind a public key to the target or ensuring freshness via a nonce.
\end{itemize}

The following deficiencies exist:
\begin{itemize}
    \item No unique identity of the target is established. Due to that, impersonation attacks cannot be detected.
    \item The measurements of the guest firmware cannot be reproduced, as it is closed source. 
    \item It is not possible to collect runtime metrics.
\end{itemize}

\subsubsection{Intel TDX CVMs}

\begin{figure}[tb]
    \centering
    \includegraphics[width=0.9\columnwidth]{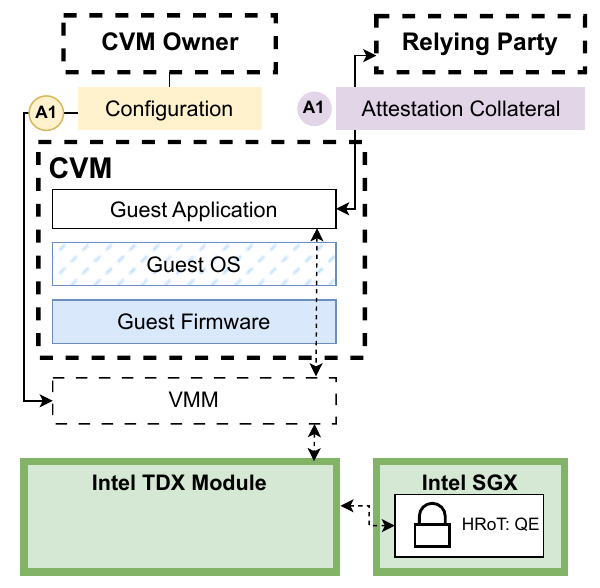}
    \caption{\ac{GCP} \ac{TDX} \ac{CVM}: Architecture.}
    \label{fig:google-tdx-cvm}
\end{figure}

Figure~\ref{fig:google-tdx-cvm} depicts the configuration and attestation of a \ac{CVM}.
Prior to launch, the owner of a \ac{CVM} can select the used guest image.
This guest image can be either a preconfigured one, or a customer provided~\cite{GoogleImage}, with a specific kernel configuration supporting \ac{TDX}.
Furthermore, similar to the offering for AMD \ac{SNP} \acp{CVM}, secure boot, vTPM, and Integrity Monitoring can be enabled as well.
The guest can retrieve quotes directly from hardware and the verification can be done using open source tools.

\paragraph{Identity Establishment}

The hardware identity used for signing the evidence is the \ac{QE} running on Intel \ac{SGX}.
The recording of loading time measurements is conducted by the \ac{TDX} module, while the recording of runtime measurements is done by the virtual firmware.
The latter allows unique identification of the target by measurement.

\paragraph{Attestation-collateral Metrics}

The loading time measurement covers the virtual firmware of the \ac{CVM}. 
The virtual firmware is not attestable, as the measurement can only be compared to a Google-provided value and not reproduced~\cite{googleFirmwareTDX}.
At runtime, the \acp{RTMR} can be used for collecting measurements.
The registers initially contain measurements of the configuration of the \ac{TD} virtual firmware, the \ac{TD} OS loader and kernel, and the OS application. 
As described, the virtual firmware is recording the measurements.
As this component is not attestable, implicit trust is required on conducting runtime measurements.
If the owner enabled secure boot, this is reflected in the event log, but not on the attestation report.

\paragraph{Trusted Computing Base}
The \ac{TCB} is composed of the Intel TD firmware, guest firmware, and guest OS.
Since the guest firmware cannot be attested and is provided by the Google, the \ac{CSP} becomes a trusted stakeholder for attestation.
Customising the \ac{TCB} is providing the OS image.

\paragraph{Conclusion}

The \ac{TDX} \ac{CVM} is able to provide the following:
\begin{itemize}
    \item The guest is able to directly retrieve \ac{TD} quotes.
    \item The offering allows the unique identification of the target via its measurements.
    \item Runtime measurements can be conducted by trusting the virtual firmware.
\end{itemize}

We notice the following limitations:
\begin{itemize}
    \item The virtual firmware is closed-source software owned by Google, therefore not attestable. Due to that, Google becomes a trusted stakeholder.
    \item Runtime measurements are collected by the virtual firmware, which cannot be verified.
\end{itemize}

\subsection{Comparison}
\label{sec:eva}

\begin{table*}[tb]
    \centering
    \begin{tabular}{l|l|cc|c|cc}
        \toprule
        \multicolumn{2}{l|}{} & \multicolumn{2}{c|}{\textbf{Microsoft Azure}} & \multicolumn{1}{c|}{\textbf{AWS}} & \multicolumn{2}{c}{\textbf{Google Cloud}} \\
         \multicolumn{2}{l|}{} & \textbf{CVM} & \textbf{Container} & \textbf{CVM} & \textbf{CVM} & \textbf{CVM}  \\
        \midrule
        \multicolumn{2}{l|}{\textbf{Platform}} & \ac{SNP} & \ac{SNP} & \ac{SNP} & \ac{SNP} & \ac{TDX} \\
        \multicolumn{2}{l|}{\textbf{Direct Retrieval of Reports}} & \FilledCircle{0} & \FilledCircle{1} & \FilledCircle{1} & \FilledCircle{1} & \FilledCircle{1}  \\
        \midrule
        \textbf{Unique} & Loading-time & \FilledCircle{0} & \FilledCircle{0} & \FilledCircle{0} & \FilledCircle{0} & \FilledCircle{1} \\
        \textbf{Target Identity} & Runtime & \FilledCircle{0.5} & \FilledCircle{1} & \FilledCircle{1} & \FilledCircle{1} & \FilledCircle{1} \\
        \midrule
        \textbf{Attestable}& Policies & \FilledCircle{0.5} & \FilledCircle{0} & \FilledCircle{0} & \FilledCircle{0} & \FilledCircle{0}  \\
        \textbf{Metrics} & Loading-time & \FilledCircle{0} & \FilledCircle{0} & \FilledCircle{1} & \FilledCircle{0} & \FilledCircle{0} \\
         & Runtime & \FilledCircle{0.5} & \FilledCircle{0} & \FilledCircle{0} & \FilledCircle{0} & \FilledCircle{0.5}  \\
        
        \midrule
        \textbf{\ac{TCB}} & Customisability & \FilledCircle{0} & \FilledCircle{0} & \FilledCircle{0.5} & \FilledCircle{0} & \FilledCircle{0} \\
        & Stakeholders & AMD, Azure & AMD, Azure & AMD, \ac{AWS} & AMD, Google & Intel, Google \\
        \bottomrule
    \end{tabular}
        \captionsetup{justification=centering} 
    \caption*{\normalfont Legend: \FilledCircle{1} supported \FilledCircle{0.5} supported with extended trust \FilledCircle{0} not supported}
    \caption{Comparison of attestation in confidential computing offerings based on the developed taxonomy.}
    \label{tab:con}
\end{table*}

Our analysis covers four hardware-based \ac{CVM} offerings, three of which are based on AMD \ac{SNP} (\ac{AWS}, Azure, \ac{GCP}) and one on Intel \ac{TDX} (\ac{GCP}).
Furthermore, we studied one confidential container offering, provided by Azure and based on AMD \ac{SNP} \acp{CVM}.
Table~\ref{tab:con} shows the results of our comparison, which is described below.


\paragraph{Identity Establishment}

An unique identity can be established during loading time by a key representing the owner or by cryptographic measurements that uniquely represent the target.
If this is not possible, a user is unable to detect impersonation attacks and therefore cannot be sure that they use the \ac{TEE} they received the attestation collateral for.
Only the \ac{TDX} \ac{CVM} offering by \ac{GCP} allows for the unique establishment of a target identity 
and this is facilitated by the representative measurement recorded in the attestation collateral.
During runtime, the target's public key can be linked to an attestation report, allowing the establishment of a secure channel with the relying party.
This can be done in all offerings, but in the case of Azure \ac{SNP} \acp{CVM} the provided verifier has to be trusted since the attestation reports cannot be retrieved directly.

\paragraph{Attestation-collateral Metrics}
Configuring security policies, such as secure boot, is possible in all solutions.
However, the enforcement of policies is attestable only in Azure \ac{SNP} \acp{CVM}, and only by depending on the provided verifier (i.e. \ac{MAA}) and the \ac{HCL}, which emulates the vTPM.
All offerings allow the collection of load-time metrics, but in most cases, they are not attestable.
Only \ac{SNP} \acp{CVM} in \ac{AWS} use open source firmware, allowing to reproduce measurements.
The collection of runtime metrics is promised by all \ac{CVM} offerings, but the recording of them needs to be performed by a verifiable RoT.
As described previously, no offering enables such verifiability.
Besides that, the RoT for collecting evidence must reside inside of the \ac{CVM}, instead of being managed by the \ac{CSP}.
This condition is fulfilled by Azure \ac{SNP} \acp{CVM} and \ac{GCP} \ac{TDX}, because of which we evaluate their attestable runtime metrics as partially supported. 

\paragraph{\acl{TCB}}
None of the solutions allows to attest all components of the \ac{TCB}, as depicted previously.
Configuring the \ac{TCB} is limited to choosing the target OS or the container image for the \ac{CVM} and container offerings respectively. 
\ac{AWS} is the only offering that allows the owner to configure secure boot with trusted signatures, so that the \ac{TCB} can be customised, but a choice over the virtual firmware is not possible.

Our trust model assumes that only the hardware vendor needs to be trusted on attestation.
None of the offerings is able to fulfil this requirement, as either closed source software, provided by the \ac{CSP}, is used, or not all components of the \ac{TCB} can be measured and therefore verified. 

\paragraph{Preliminary Conclusion}
We see a strong discrepancy between the security principles needed for confidential computing and what is actually being offered in commercial solutions.
None of the studied architectures provides all the necessary building blocks we expect for attestation, leading to the \ac{CSP} becoming an implicitly trusted stakeholder.
Strictly speaking, the use of these solutions is rendered superfluous, as it is of the utmost importance to keep the infrastructure provider outside of the trust model and comply with the confidential computing paradigm.

\subsection{Identification of Root Causes}
Despite the promise of confidential computing to provide hardware-backed security guarantees for workloads running on untrusted cloud infrastructures, based on our findings, public cloud offerings are not capable of achieving this vision. 
Although the use of \acp{TEE} creates expectations of minimised trust dependencies, in practice, cloud providers retain significant control over key components of the system and its attestation.
This discrepancy arises from various factors, including legal obligations, operational constraints, and economic incentives, which ultimately influence how the architectures of the offered solutions are being designed. 

When it comes to legal obligations, government entities may exert pressure on cloud providers to grant access to customer data. 
For instance, under legal frameworks like the U.S. CLOUD Act~\cite{CLOUDAct2018}, service providers can be compelled to disclose data stored on their servers, regardless of the data's physical location. 
This is also applicable to public sector organisations that often operate under stringent compliance requirements concerning data sovereignty and privacy. 
While intended to enhance security and compliance, arrangements set to meet these requirements can introduce additional trust dependencies that deviate from the pure confidential computing paradigm. 
The shortfalls we have identified in this paper might also be rooted in some operational constraints where part of the \ac{TCB} within the TEE, i.e. virtual firmware, cannot be provided by cloud customers because of technical reasons or lack of resources and necessarily falls under the responsibility of the cloud provider. 
This unavoidably comes with some additional caveats, including transparency of the code used, centralised control over it, and lack of flexibility towards the cloud customer. 

Last but not least, in some use-case scenarios where multiple parties engage in collaborative computations, like anti-money laundering, collaborative drug development~\cite{AzureConfidentialComputing} or privacy-preserving machine learning~\cite{Bogdanov2018}, the primary concern often centres on protecting each entity's proprietary data and code from other participants, rather than focussing on the cloud provider's access. 
In such cases, the cloud provider is often considered a neutral party, and the emphasis is on ensuring that the collaborating entities cannot access each other's sensitive information.
Therefore, even though the parties involved in these computations are deeply concerned about certain security guarantees, their trust model does not necessarily exclude the cloud provider. 

\section{Conclusion}
\label{sec:con}

In this paper, we examined the confidential computing offerings of the three major \acp{CSP}, namely Azure, \ac{AWS}, and \ac{GCP}, focussing on \ac{VM}-based execution contexts using AMD \ac{SNP} and Intel \ac{TDX}.
Leveraging our proposed taxonomy, we analysed these solutions in terms of \ac{CVM} setup and remote attestation.
Our analysis revealed significant gaps between the intended threat model and the realities of current implementations. 
In particular, the \acp{CSP} retain control over critical components of the \ac{TCB}, including those responsible for remote attestation. 
Moreover, entities tasked with recording and reporting the \ac{TCB} state are either unverifiable or exist outside the \ac{TEE}, remaining under \ac{CSP} control.  
Given that remote attestation is the cornerstone of trust in confidential computing, it is crucial to emphasise that the way cloud providers market these solutions may give customers a misleading sense of security regarding the confidentiality and integrity of their data. 

\section*{Acknowledgement}
This work has been partly funded by the German Federal Office for Information Security (BSI) project 6G-ReS (grant no. 01MO23013D), the Federal Ministry of Education and Research of Germany project 6G-CampuSens (project id. 16KISK205) as well as SEMECO-Q1 (grant no. 03ZU1210AA) and IPCEI-CIS (grant no. 13IPC005). Additionally, the authors are also financed based on the budget passed by the Saxonian State Parliament in Germany.  

\bibliography{bibliography}{}


\begin{thebibliography}{61}


\ifx \showCODEN    \undefined \def \showCODEN     #1{\unskip}     \fi
\ifx \showDOI      \undefined \def \showDOI       #1{#1}\fi
\ifx \showISBNx    \undefined \def \showISBNx     #1{\unskip}     \fi
\ifx \showISBNxiii \undefined \def \showISBNxiii  #1{\unskip}     \fi
\ifx \showISSN     \undefined \def \showISSN      #1{\unskip}     \fi
\ifx \showLCCN     \undefined \def \showLCCN      #1{\unskip}     \fi
\ifx \shownote     \undefined \def \shownote      #1{#1}          \fi
\ifx \showarticletitle \undefined \def \showarticletitle #1{#1}   \fi
\ifx \showURL      \undefined \def \showURL       {\relax}        \fi
\providecommand\bibfield[2]{#2}
\providecommand\bibinfo[2]{#2}
\providecommand\natexlab[1]{#1}
\providecommand\showeprint[2][]{arXiv:#2}

\bibitem[CLO(2018)]%
        {CLOUDAct2018}
 \bibinfo{year}{2018}\natexlab{}.
\newblock \bibinfo{title}{CLOUD Act}.
\newblock \bibinfo{howpublished}{\url{https://www.justice.gov/d9/pages/attachments/2019/04/09/cloud\_act.pdf}}.
\newblock


\bibitem[{Advanced Micro Devices, Inc.}(2023a)]%
        {AMDSVSM}
\bibfield{author}{\bibinfo{person}{{Advanced Micro Devices, Inc.}}} \bibinfo{year}{2023}\natexlab{a}.
\newblock \bibinfo{title}{Secure VM Service Module for SEV-SNP Guests}.
\newblock
\newblock
\urldef\tempurl%
\url{https://www.amd.com/content/dam/amd/en/documents/epyc-technical-docs/specifications/58019.pdf}
\showURL{%
\tempurl}


\bibitem[{Advanced Micro Devices, Inc.}(2023b)]%
        {AMDSEVSNPABI}
\bibfield{author}{\bibinfo{person}{{Advanced Micro Devices, Inc.}}} \bibinfo{year}{2023}\natexlab{b}.
\newblock \bibinfo{booktitle}{\emph{SEV Secure Nested Paging Firmware ABI Specification}}.
\newblock \bibinfo{type}{Technical Specification} 56860. \bibinfo{institution}{AMD}.
\newblock
\urldef\tempurl%
\url{https://www.amd.com/content/dam/amd/en/documents/epyc-technical-docs/specifications/56860.pdf}
\showURL{%
\tempurl}
\newblock
\shownote{Accessed on February 02, 2025}.


\bibitem[{Advanced Micro Devices, Inc.}(2025)]%
        {AMDSEVSNP}
\bibfield{author}{\bibinfo{person}{{Advanced Micro Devices, Inc.}}} \bibinfo{year}{2025}\natexlab{}.
\newblock \bibinfo{booktitle}{\emph{AMD SEV-SNP: Strengthening VM Isolation with Integrity Protection and More}}.
\newblock \bibinfo{type}{White Paper}. \bibinfo{institution}{AMD}.
\newblock
\newblock
\shownote{Accessed on February 02, 2025}.


\bibitem[{Amazon Web Services, Inc.}(2021)]%
        {awsConfComp}
\bibfield{author}{\bibinfo{person}{{Amazon Web Services, Inc.}}} \bibinfo{year}{2021}\natexlab{}.
\newblock \bibinfo{title}{{Confidential computing: an AWS perspective {$\vert$} Amazon Web Services}}.
\newblock \bibinfo{howpublished}{\url{https://aws.amazon.com/blogs/security/confidential-computing-an-aws-perspective}}.
\newblock
\newblock
\shownote{[Online; accessed 28. Feb. 2025]}.


\bibitem[{Amazon Web Services, Inc}(2025a)]%
        {AmazonNitro}
\bibfield{author}{\bibinfo{person}{{Amazon Web Services, Inc}}.} \bibinfo{year}{2025}\natexlab{a}.
\newblock \bibinfo{title}{{Nitro Enclaves}}.
\newblock \bibinfo{howpublished}{\url{https://aws.amazon.com/ec2/nitro/nitro-enclaves}}.
\newblock
\newblock
\shownote{[Online; accessed 6. Feb. 2025]}.


\bibitem[{Amazon Web Services, Inc}(2025b)]%
        {NitroTPM}
\bibfield{author}{\bibinfo{person}{{Amazon Web Services, Inc}}.} \bibinfo{year}{2025}\natexlab{b}.
\newblock \bibinfo{title}{{NitroTPM for Amazon EC2 instances - Amazon Elastic Compute Cloud}}.
\newblock \bibinfo{howpublished}{\url{https://docs.aws.amazon.com/AWSEC2/latest/UserGuide/nitrotpm.html}}.
\newblock
\newblock
\shownote{[Online; accessed 11. Feb. 2025]}.


\bibitem[{Amazon Web Services, Inc}(2025c)]%
        {awsUefi}
\bibfield{author}{\bibinfo{person}{{Amazon Web Services, Inc}}.} \bibinfo{year}{2025}\natexlab{c}.
\newblock \bibinfo{title}{{uefi}}.
\newblock \bibinfo{howpublished}{\url{https://github.com/aws/uefi}}.
\newblock
\newblock
\shownote{[Online; accessed 6. Feb. 2025]}.


\bibitem[{Amazon Web Services, Inc}(2025d)]%
        {awsuefipatch}
\bibfield{author}{\bibinfo{person}{{Amazon Web Services, Inc}}.} \bibinfo{year}{2025}\natexlab{d}.
\newblock \bibinfo{title}{{uefi}}.
\newblock \bibinfo{howpublished}{\url{https://github.com/aws/uefi/blob/5c3ac896feea3923a96944dc23e808385939c0a1/edk2-stable202211/0026-edk2-stable202211-OvmfPkg-PlatformPei-Initialize-variable-store-after-SEV.patch\#L4}}.
\newblock
\newblock
\shownote{[Online; accessed 20. Feb. 2025]}.


\bibitem[Arnautov et~al\mbox{.}(2016)]%
        {arnautov16scone}
\bibfield{author}{\bibinfo{person}{Sergei Arnautov}, \bibinfo{person}{Bohdan Trach}, \bibinfo{person}{Franz Gregor}, \bibinfo{person}{Thomas Knauth}, \bibinfo{person}{Andre Martin}, \bibinfo{person}{Christian Priebe}, \bibinfo{person}{Joshua Lind}, \bibinfo{person}{Divya Muthukumaran}, \bibinfo{person}{Dan O{\textquoteright}Keeffe}, \bibinfo{person}{Mark~L. Stillwell}, \bibinfo{person}{David Goltzsche}, \bibinfo{person}{Dave Eyers}, \bibinfo{person}{R{\"u}diger Kapitza}, \bibinfo{person}{Peter Pietzuch}, {and} \bibinfo{person}{Christof Fetzer}.} \bibinfo{year}{2016}\natexlab{}.
\newblock \showarticletitle{{SCONE}: Secure Linux Containers with Intel {SGX}}. In \bibinfo{booktitle}{\emph{12th USENIX Symposium on Operating Systems Design and Implementation (OSDI 16)}}. \bibinfo{publisher}{USENIX Association}, \bibinfo{address}{Savannah, GA}, \bibinfo{pages}{689--703}.
\newblock
\showISBNx{978-1-931971-33-1}
\urldef\tempurl%
\url{https://www.usenix.org/conference/osdi16/technical-sessions/presentation/arnautov}
\showURL{%
\tempurl}


\bibitem[Baumann et~al\mbox{.}(2015)]%
        {baumann15haven}
\bibfield{author}{\bibinfo{person}{Andrew Baumann}, \bibinfo{person}{Marcus Peinado}, {and} \bibinfo{person}{Galen Hunt}.} \bibinfo{year}{2015}\natexlab{}.
\newblock \showarticletitle{Shielding Applications from an Untrusted Cloud with Haven}.
\newblock \bibinfo{journal}{\emph{ACM Trans. Comput. Syst.}} \bibinfo{volume}{33}, \bibinfo{number}{3}, Article \bibinfo{articleno}{8} (\bibinfo{date}{Aug.} \bibinfo{year}{2015}), \bibinfo{numpages}{26}~pages.
\newblock
\showISSN{0734-2071}
\urldef\tempurl%
\url{https://doi.org/10.1145/2799647}
\showDOI{\tempurl}


\bibitem[Berger et~al\mbox{.}(2006)]%
        {Berger2006vTPMVT}
\bibfield{author}{\bibinfo{person}{Stefan Berger}, \bibinfo{person}{Ram{'o}n C{'a}ceres}, \bibinfo{person}{Kenneth~A. Goldman}, \bibinfo{person}{Ronald Perez}, \bibinfo{person}{Reiner Sailer}, {and} \bibinfo{person}{Leendert van Doorn}.} \bibinfo{year}{2006}\natexlab{}.
\newblock \showarticletitle{vTPM: Virtualizing the Trusted Platform Module}. In \bibinfo{booktitle}{\emph{USENIX Security Symposium}}.
\newblock
\urldef\tempurl%
\url{https://api.semanticscholar.org/CorpusID:17899967}
\showURL{%
\tempurl}


\bibitem[Birkholz et~al\mbox{.}(2023)]%
        {RFC9334}
\bibfield{author}{\bibinfo{person}{Henk Birkholz}, \bibinfo{person}{Dave Thaler}, \bibinfo{person}{Michael Richardson}, \bibinfo{person}{Ned Smith}, {and} \bibinfo{person}{Wei Pan}.} \bibinfo{year}{2023}\natexlab{}.
\newblock \bibinfo{title}{{Remote ATtestation procedureS (RATS) Architecture}}.
\newblock \bibinfo{howpublished}{RFC 9334}.
\newblock
\urldef\tempurl%
\url{https://doi.org/10.17487/RFC9334}
\showDOI{\tempurl}


\bibitem[Bogdanov et~al\mbox{.}(2018)]%
        {Bogdanov2018}
\bibfield{author}{\bibinfo{person}{Dan Bogdanov}, \bibinfo{person}{Sven Laur}, {and} \bibinfo{person}{Jan Willemson}.} \bibinfo{year}{2018}\natexlab{}.
\newblock \showarticletitle{Real-World Applications of Secure Multi-Party Computation}.
\newblock \bibinfo{journal}{\emph{Cryptology ePrint Archive}} (\bibinfo{year}{2018}).
\newblock
\urldef\tempurl%
\url{https://eprint.iacr.org/2018/450.pdf}
\showURL{%
\tempurl}


\bibitem[Cheng et~al\mbox{.}(2024)]%
        {tdxdemystified}
\bibfield{author}{\bibinfo{person}{Pau-Chen Cheng}, \bibinfo{person}{Wojciech Ozga}, \bibinfo{person}{Enriquillo Valdez}, \bibinfo{person}{Salman Ahmed}, \bibinfo{person}{Zhongshu Gu}, \bibinfo{person}{Hani Jamjoom}, \bibinfo{person}{Hubertus Franke}, {and} \bibinfo{person}{James Bottomley}.} \bibinfo{year}{2024}\natexlab{}.
\newblock \showarticletitle{Intel TDX Demystified: A Top-Down Approach}.
\newblock \bibinfo{journal}{\emph{ACM Comput. Surv.}} \bibinfo{volume}{56}, \bibinfo{number}{9}, Article \bibinfo{articleno}{238} (\bibinfo{date}{April} \bibinfo{year}{2024}), \bibinfo{numpages}{33}~pages.
\newblock
\showISSN{0360-0300}
\urldef\tempurl%
\url{https://doi.org/10.1145/3652597}
\showDOI{\tempurl}


\bibitem[Cloud(2025a)]%
        {googleMonitoring}
\bibfield{author}{\bibinfo{person}{Google Cloud}.} \bibinfo{year}{2025}\natexlab{a}.
\newblock \bibinfo{title}{{Monitoring integrity on Shielded VMs}}.
\newblock \bibinfo{howpublished}{\url{https://cloud.google.com/compute/shielded-vm/docs/integrity-monitoring}}.
\newblock
\newblock
\shownote{[Online; accessed 11. Feb. 2025]}.


\bibitem[Cloud(2025b)]%
        {googleFirmware}
\bibfield{author}{\bibinfo{person}{Google Cloud}.} \bibinfo{year}{2025}\natexlab{b}.
\newblock \bibinfo{title}{{Verify a Confidential VM instance's firmware}}.
\newblock \bibinfo{howpublished}{\url{https://cloud.google.com/confidential-computing/confidential-vm/docs/verify-firmware}}.
\newblock
\newblock
\shownote{[Online; accessed 11. Feb. 2025]}.


\bibitem[Cloud(2025c)]%
        {googleShieldedVM}
\bibfield{author}{\bibinfo{person}{Google Cloud}.} \bibinfo{year}{2025}\natexlab{c}.
\newblock \bibinfo{title}{{What is Shielded VM?}}
\newblock \bibinfo{howpublished}{\url{https://cloud.google.com/compute/shielded-vm/docs/shielded-vm}}.
\newblock
\newblock
\shownote{[Online; accessed 11. Feb. 2025]}.


\bibitem[Corporation(2014)]%
        {sgxreference}
\bibfield{author}{\bibinfo{person}{Intel Corporation}.} \bibinfo{year}{2014}\natexlab{}.
\newblock \bibinfo{booktitle}{\emph{Intel Software Guard Extensions Programming Reference}}.
\newblock \bibinfo{type}{{T}echnical {R}eport} 329298-002US. \bibinfo{institution}{Intel Corporation}.
\newblock
\urldef\tempurl%
\url{https://www.intel.com/content/dam/develop/external/us/en/documents/329298-002-629101.pdf}
\showURL{%
\tempurl}


\bibitem[Corporation(2023)]%
        {IntelSGXDCAP2023}
\bibfield{author}{\bibinfo{person}{Intel Corporation}.} \bibinfo{year}{2023}\natexlab{}.
\newblock \bibinfo{booktitle}{\emph{Intel\textregistered{} SGX Data Center Attestation Primitives (Intel\textregistered{} SGX DCAP)}}.
\newblock \bibinfo{type}{{T}echnical {R}eport} 344991-004US. \bibinfo{institution}{Intel Corporation}.
\newblock
\urldef\tempurl%
\url{https://www.intel.com/content/dam/develop/public/us/en/documents/intel-sgx-dcap-ecdsa-orientation.pdf}
\showURL{%
\tempurl}


\bibitem[{Decentriq}(2024)]%
        {DecentriqSEVSNP}
\bibfield{author}{\bibinfo{person}{{Decentriq}}.} \bibinfo{year}{2024}\natexlab{}.
\newblock \bibinfo{title}{Swiss cheese to cheddar: securing AMD SEV-SNP early boot}.
\newblock
\newblock
\urldef\tempurl%
\url{https://www.decentriq.com/article/swiss-cheese-to-cheddar-securing-amd-sev-snp-early-boot}
\showURL{%
\tempurl}
\newblock
\shownote{Accessed on February 02, 2025}.


\bibitem[{Enclaive}(2025)]%
        {EnclaiveVMPL}
\bibfield{author}{\bibinfo{person}{{Enclaive}}.} \bibinfo{year}{2025}\natexlab{}.
\newblock \bibinfo{title}{Virtual Machine Privilege Levels | Confidential Computing 101}.
\newblock \bibinfo{howpublished}{\url{https://docs.enclaive.cloud/confidential-cloud/technology-in-depth/amd-sev/technology/fundamentals/features/virtual-machine-privilege-levels}}.
\newblock
\newblock
\shownote{Accessed on February 02, 2025}.


\bibitem[Galanou et~al\mbox{.}(2023)]%
        {revelio}
\bibfield{author}{\bibinfo{person}{Anna Galanou}, \bibinfo{person}{Khushboo Bindlish}, \bibinfo{person}{Luca Preibsch}, \bibinfo{person}{Yvonne-Anne Pignolet}, \bibinfo{person}{Christof Fetzer}, {and} \bibinfo{person}{Rüdiger Kapitza}.} \bibinfo{year}{2023}\natexlab{}.
\newblock \showarticletitle{Trustworthy confidential virtual machines for the masses}. In \bibinfo{booktitle}{\emph{Middleware 2023 - Proceedings of the 24th ACM/IFIP International Middleware Conference}}. \bibinfo{pages}{316--328}.
\newblock
\showISBNx{9798400701771}
\urldef\tempurl%
\url{https://doi.org/10.1145/3590140.3629124}
\showDOI{\tempurl}


\bibitem[Ge et~al\mbox{.}(2022)]%
        {ge2022hecate}
\bibfield{author}{\bibinfo{person}{Xinyang Ge}, \bibinfo{person}{Hsuan-Chi Kuo}, {and} \bibinfo{person}{Weidong Cui}.} \bibinfo{year}{2022}\natexlab{}.
\newblock \showarticletitle{Hecate: Lifting and shifting on-premises workloads to an untrusted cloud}. In \bibinfo{booktitle}{\emph{Proceedings of the 2022 ACM SIGSAC Conference on Computer and Communications Security}}. \bibinfo{pages}{1231--1242}.
\newblock


\bibitem[{gematik}(2024a)]%
        {gem2}
\bibfield{author}{\bibinfo{person}{{gematik}}.} \bibinfo{year}{2024}\natexlab{a}.
\newblock \bibinfo{title}{{E-Rezept\_Feature-Dokumente}}.
\newblock \bibinfo{howpublished}{\url{https://gemspec.gematik.de/releases/E-Rezept\_Feature-Dokumente/}}.
\newblock
\newblock
\shownote{[Online; accessed 27. Feb. 2025]}.


\bibitem[{gematik}(2024b)]%
        {gem3}
\bibfield{author}{\bibinfo{person}{{gematik}}.} \bibinfo{year}{2024}\natexlab{b}.
\newblock \bibinfo{title}{{Prerelease: Dev\_HCC}}.
\newblock \bibinfo{howpublished}{\url{https://gemspec.gematik.de/prereleases/Dev\_HCC/}}.
\newblock
\newblock
\shownote{[Online; accessed 27. Feb. 2025]}.


\bibitem[{gematik}(2025)]%
        {gem1}
\bibfield{author}{\bibinfo{person}{{gematik}}.} \bibinfo{year}{2025}\natexlab{}.
\newblock \bibinfo{title}{{Spezifikation Sektoraler Identity Provider}}.
\newblock \bibinfo{howpublished}{\url{https://gemspec.gematik.de/docs/gemSpec/gemSpec\_IDP\_Sek/latest/}}.
\newblock
\newblock
\shownote{[Online; accessed 27. Feb. 2025]}.


\bibitem[{Google Cloud}(2025a)]%
        {GoogleConfComp}
\bibfield{author}{\bibinfo{person}{{Google Cloud}}.} \bibinfo{year}{2025}\natexlab{a}.
\newblock \bibinfo{title}{{Confidential Computing {$\vert$} Google Cloud}}.
\newblock \bibinfo{howpublished}{\url{https://cloud.google.com/security/products/confidential-computing?hl=en}}.
\newblock
\newblock
\shownote{[Online; accessed 28. Feb. 2025]}.


\bibitem[{Google Cloud}(2025b)]%
        {GoogleImage}
\bibfield{author}{\bibinfo{person}{{Google Cloud}}.} \bibinfo{year}{2025}\natexlab{b}.
\newblock \bibinfo{title}{{Create custom Confidential VM images}}.
\newblock \bibinfo{howpublished}{\url{https://cloud.google.com/confidential-computing/confidential-vm/docs/create-custom-confidential-vm-images}}.
\newblock
\newblock
\shownote{[Online; accessed 10. Feb. 2025]}.


\bibitem[{Google Cloud}(2025c)]%
        {googleFirmwareTDX}
\bibfield{author}{\bibinfo{person}{{Google Cloud}}.} \bibinfo{year}{2025}\natexlab{c}.
\newblock \bibinfo{title}{{Verify a Confidential VM instance's firmware}}.
\newblock \bibinfo{howpublished}{\url{https://cloud.google.com/confidential-computing/confidential-vm/docs/verify-firmware\#intel-tdx}}.
\newblock
\newblock
\shownote{[Online; accessed 25. Feb. 2025]}.


\bibitem[{Intel Corporation}(2023)]%
        {IntelTDX2023}
\bibfield{author}{\bibinfo{person}{{Intel Corporation}}.} \bibinfo{year}{2023}\natexlab{}.
\newblock \bibinfo{booktitle}{\emph{{Intel\textregistered{} TDX Virtual Firmware Design Guide}}}.
\newblock \bibinfo{type}{{T}echnical {R}eport} 344991-004US. \bibinfo{institution}{Intel Corporation}.
\newblock
\urldef\tempurl%
\url{https://cdrdv2-public.intel.com/733585/tdx-virtual-firmware-design-guide-rev-004-20231206.pdf}
\showURL{%
\tempurl}


\bibitem[{Intel Corporation}(2024a)]%
        {IntelTDXOverview}
\bibfield{author}{\bibinfo{person}{{Intel Corporation}}.} \bibinfo{year}{2024}\natexlab{a}.
\newblock \bibinfo{title}{Intel® Trust Domain Extensions (Intel® TDX)}.
\newblock \bibinfo{howpublished}{\url{https://www.intel.com/content/www/us/en/developer/tools/trust-domain-extensions/overview.html}}.
\newblock
\newblock
\shownote{Accessed on February 02, 2025}.


\bibitem[{Intel Corporation}(2024b)]%
        {IntelTDXDocumentation}
\bibfield{author}{\bibinfo{person}{{Intel Corporation}}.} \bibinfo{year}{2024}\natexlab{b}.
\newblock \bibinfo{title}{Intel® Trust Domain Extensions (Intel® TDX) - Documentation}.
\newblock \bibinfo{howpublished}{\url{https://www.intel.com/content/www/us/en/developer/tools/trust-domain-extensions/documentation.html}}.
\newblock
\newblock
\shownote{Accessed on February 02, 2025}.


\bibitem[{Intel Corporation}(2025)]%
        {IntelTDXGuide}
\bibfield{author}{\bibinfo{person}{{Intel Corporation}}.} \bibinfo{year}{2025}\natexlab{}.
\newblock \bibinfo{title}{Intel TDX Enabling Guide}.
\newblock \bibinfo{howpublished}{\url{https://cc-enabling.trustedservices.intel.com/intel-tdx-enabling-guide/01/introduction/}}.
\newblock
\urldef\tempurl%
\url{https://cc-enabling.trustedservices.intel.com/intel-tdx-enabling-guide/01/introduction/}
\showURL{%
\tempurl}
\newblock
\shownote{Accessed on February 02, 2025}.


\bibitem[Johnson et~al\mbox{.}(2024)]%
        {johnson2024confidential}
\bibfield{author}{\bibinfo{person}{Matthew~A Johnson}, \bibinfo{person}{Stavros Volos}, \bibinfo{person}{Ken Gordon}, \bibinfo{person}{Sean~T Allen}, \bibinfo{person}{Christoph~M Wintersteiger}, \bibinfo{person}{Sylvan Clebsch}, \bibinfo{person}{John Starks}, {and} \bibinfo{person}{Manuel Costa}.} \bibinfo{year}{2024}\natexlab{}.
\newblock \showarticletitle{Confidential Container Groups: Implementing confidential computing on Azure container instances}.
\newblock \bibinfo{journal}{\emph{Queue}} \bibinfo{volume}{22}, \bibinfo{number}{2} (\bibinfo{year}{2024}), \bibinfo{pages}{57--86}.
\newblock


\bibitem[{Juniper Networks}(2020)]%
        {juniper2020securemeasuredboot}
\bibfield{author}{\bibinfo{person}{{Juniper Networks}}.} \bibinfo{year}{2020}\natexlab{}.
\newblock \bibinfo{title}{What's the Difference between Secure Boot and Measured Boot?}
\newblock
\newblock
\urldef\tempurl%
\url{https://community.juniper.net/blogs/elevate-member/2020/12/22/whats-the-difference-between-secure-boot-and-measured-boot}
\showURL{%
\tempurl}


\bibitem[kernel~development community(2025)]%
        {dmverity}
\bibfield{author}{\bibinfo{person}{The kernel~development community}.} \bibinfo{year}{2025}\natexlab{}.
\newblock \bibinfo{title}{{dm-verity}}.
\newblock \bibinfo{howpublished}{\url{https://docs.kernel.org/admin-guide/device-mapper/verity.html}}.
\newblock
\newblock
\shownote{[Online; accessed 21. Feb. 2025]}.


\bibitem[Khan et~al\mbox{.}(2021)]%
        {CCforSensing}
\bibfield{author}{\bibinfo{person}{Arslan Khan}, \bibinfo{person}{Joseph~I. Choi}, \bibinfo{person}{Dave~Jing Tian}, \bibinfo{person}{Tyler Ward}, \bibinfo{person}{Kevin R.~B. Butler}, \bibinfo{person}{Patrick Traynor}, \bibinfo{person}{John~M. Shea}, {and} \bibinfo{person}{Tan~F. Wong}.} \bibinfo{year}{2021}\natexlab{}.
\newblock \showarticletitle{Privacy-{Preserving} {Localization} using {Enclaves}}. In \bibinfo{booktitle}{\emph{2021 {IEEE} 12th {Annual} {Ubiquitous} {Computing}, {Electronics} \& {Mobile} {Communication} {Conference} ({UEMCON})}}. \bibinfo{pages}{0269--0278}.
\newblock
\urldef\tempurl%
\url{https://doi.org/10.1109/UEMCON53757.2021.9666706}
\showDOI{\tempurl}


\bibitem[Kollenda(2022)]%
        {KollendaSEVTDX}
\bibfield{author}{\bibinfo{person}{Kevin Kollenda}.} \bibinfo{year}{2022}\natexlab{}.
\newblock \bibinfo{booktitle}{\emph{General overview of AMD SEV-SNP and Intel TDX}}.
\newblock \bibinfo{type}{Technical Report}. \bibinfo{institution}{Friedrich-Alexander-Universität Erlangen-Nürnberg}.
\newblock
\urldef\tempurl%
\url{https://sys.cs.fau.de/extern/lehre/ws22/akss/material/amd-sev-intel-tdx.pdf}
\showURL{%
\tempurl}
\newblock
\shownote{Accessed on February 02, 2025}.


\bibitem[Larabel(2023)]%
        {LinuxSVSM}
\bibfield{author}{\bibinfo{person}{Michael Larabel}.} \bibinfo{year}{2023}\natexlab{}.
\newblock \showarticletitle{The Linux SVSM project}.
\newblock \bibinfo{journal}{\emph{LWN.net}} (\bibinfo{date}{January} \bibinfo{year}{2023}).
\newblock
\urldef\tempurl%
\url{https://lwn.net/Articles/921266/}
\showURL{%
\tempurl}


\bibitem[Li et~al\mbox{.}(2024)]%
        {li2024sok}
\bibfield{author}{\bibinfo{person}{Mengyuan Li}, \bibinfo{person}{Yuheng Yang}, \bibinfo{person}{Guoxing Chen}, \bibinfo{person}{Mengjia Yan}, {and} \bibinfo{person}{Yinqian Zhang}.} \bibinfo{year}{2024}\natexlab{}.
\newblock \showarticletitle{Sok: Understanding design choices and pitfalls of trusted execution environments}. In \bibinfo{booktitle}{\emph{Proceedings of the 19th ACM Asia Conference on Computer and Communications Security}}. \bibinfo{pages}{1600--1616}.
\newblock


\bibitem[M{\'e}n{\'e}trey et~al\mbox{.}(2022)]%
        {menetrey2022attestation}
\bibfield{author}{\bibinfo{person}{J{\"a}mes M{\'e}n{\'e}trey}, \bibinfo{person}{Christian G{\"o}ttel}, \bibinfo{person}{Anum Khurshid}, \bibinfo{person}{Marcelo Pasin}, \bibinfo{person}{Pascal Felber}, \bibinfo{person}{Valerio Schiavoni}, {and} \bibinfo{person}{Shahid Raza}.} \bibinfo{year}{2022}\natexlab{}.
\newblock \showarticletitle{Attestation mechanisms for trusted execution environments demystified}. In \bibinfo{booktitle}{\emph{IFIP International Conference on Distributed Applications and Interoperable Systems}}. Springer, \bibinfo{pages}{95--113}.
\newblock


\bibitem[{Microsoft Azure}(2023)]%
        {AzureConfidentialComputing}
\bibfield{author}{\bibinfo{person}{{Microsoft Azure}}.} \bibinfo{year}{2023}\natexlab{}.
\newblock \bibinfo{title}{Common Azure confidential computing scenarios and use cases}.
\newblock \bibinfo{howpublished}{\url{https://learn.microsoft.com/en-us/azure/confidential-computing/use-cases-scenarios}}.
\newblock


\bibitem[{Microsoft Azure}(2024a)]%
        {azureConfContainers}
\bibfield{author}{\bibinfo{person}{{Microsoft Azure}}.} \bibinfo{year}{2024}\natexlab{a}.
\newblock \bibinfo{title}{{Confidential containers on Azure}}.
\newblock \bibinfo{howpublished}{\url{https://learn.microsoft.com/en-us/azure/confidential-computing/confidential-containers}}.
\newblock
\newblock
\shownote{[Online; accessed 12. Feb. 2025]}.


\bibitem[{Microsoft Azure}(2024b)]%
        {azurevTPMtutorial}
\bibfield{author}{\bibinfo{person}{{Microsoft Azure}}.} \bibinfo{year}{2024}\natexlab{b}.
\newblock \bibinfo{title}{{Leverage virtual TPMs in Azure confidential VMs}}.
\newblock \bibinfo{howpublished}{\url{https://learn.microsoft.com/en-us/azure/confidential-computing/how-to-leverage-virtual-tpms-in-azure-confidential-vms}}.
\newblock
\newblock
\shownote{[Online; accessed 25. Feb. 2025]}.


\bibitem[{Microsoft Azure}(2025a)]%
        {azureConfComp}
\bibfield{author}{\bibinfo{person}{{Microsoft Azure}}.} \bibinfo{year}{2025}\natexlab{a}.
\newblock \bibinfo{title}{{Azure Confidential Computing {\textendash} Protect Data In Use {$\vert$} Microsoft Azure}}.
\newblock \bibinfo{howpublished}{\url{https://azure.microsoft.com/en-us/solutions/confidential-compute}}.
\newblock
\newblock
\shownote{[Online; accessed 28. Feb. 2025]}.


\bibitem[{Microsoft Azure}(2025b)]%
        {AzurevTPM}
\bibfield{author}{\bibinfo{person}{{Microsoft Azure}}.} \bibinfo{year}{2025}\natexlab{b}.
\newblock \bibinfo{title}{{Azure Confidential VM guest attestation design detail}}.
\newblock \bibinfo{howpublished}{\url{https://learn.microsoft.com/en-us/azure/confidential-computing/guest-attestation-confidential-virtual-machines-design}}.
\newblock
\newblock
\shownote{[Online; accessed 10. Feb. 2025]}.


\bibitem[{Microsoft Azure}(2025c)]%
        {MicrosoftArm}
\bibfield{author}{\bibinfo{person}{{Microsoft Azure}}.} \bibinfo{year}{2025}\natexlab{c}.
\newblock \bibinfo{title}{{Create an Azure confidential VM with ARM template}}.
\newblock \bibinfo{howpublished}{\url{https://learn.microsoft.com/en-us/azure/confidential-computing/quick-create-confidential-vm-arm}}.
\newblock
\newblock
\shownote{[Online; accessed 6. Feb. 2025]}.


\bibitem[Mo et~al\mbox{.}(2024)]%
        {CCforML}
\bibfield{author}{\bibinfo{person}{Fan Mo}, \bibinfo{person}{Zahra Tarkhani}, {and} \bibinfo{person}{Hamed Haddadi}.} \bibinfo{year}{2024}\natexlab{}.
\newblock \showarticletitle{Machine {Learning} with {Confidential} {Computing}: {A} {Systematization} of {Knowledge}}.
\newblock \bibinfo{journal}{\emph{ACM Comput. Surv.}} \bibinfo{volume}{56}, \bibinfo{number}{11} (\bibinfo{date}{June} \bibinfo{year}{2024}), \bibinfo{pages}{281:1--281:40}.
\newblock
\showISSN{0360-0300}
\urldef\tempurl%
\url{https://doi.org/10.1145/3670007}
\showDOI{\tempurl}


\bibitem[Moriarty and Smith(2024)]%
        {RATSOverview}
\bibfield{author}{\bibinfo{person}{Kathleen Moriarty} {and} \bibinfo{person}{Ned Smith}.} \bibinfo{year}{2024}\natexlab{}.
\newblock \bibinfo{title}{Remote Attestation Procedures IETF Working Group Overview and Engagement}.
\newblock
\newblock
\urldef\tempurl%
\url{https://crypto.orange-labs.fr/acg/workshop/pdf/6-IETF-Remote_Attestation_Procedures_-_Attestation_Workshop-nms1.pdf}
\showURL{%
\tempurl}


\bibitem[Mu{\~n}oz et~al\mbox{.}(2023)]%
        {munoz2023survey}
\bibfield{author}{\bibinfo{person}{Antonio Mu{\~n}oz}, \bibinfo{person}{Ruben R{\'\i}os}, \bibinfo{person}{Rodrigo Rom{\'a}n}, {and} \bibinfo{person}{Javier L{\'o}pez}.} \bibinfo{year}{2023}\natexlab{}.
\newblock \showarticletitle{A survey on the (in) security of trusted execution environments}.
\newblock \bibinfo{journal}{\emph{Computers \& Security}}  \bibinfo{volume}{129} (\bibinfo{year}{2023}), \bibinfo{pages}{103180}.
\newblock


\bibitem[Niemi et~al\mbox{.}(2022)]%
        {niemi2022towards}
\bibfield{author}{\bibinfo{person}{Arto Niemi}, \bibinfo{person}{Sampo Sovio}, {and} \bibinfo{person}{Jan-Erik Ekberg}.} \bibinfo{year}{2022}\natexlab{}.
\newblock \showarticletitle{Towards interoperable enclave attestation: Learnings from decades of academic work}. In \bibinfo{booktitle}{\emph{2022 31st Conference of Open Innovations Association (FRUCT)}}. IEEE, \bibinfo{pages}{189--200}.
\newblock


\bibitem[Pulapaka(2024)]%
        {PulapakaOpenHCL}
\bibfield{author}{\bibinfo{person}{Hari Pulapaka}.} \bibinfo{year}{2024}\natexlab{}.
\newblock \bibinfo{title}{{OpenHCL: Evolving Azure's virtualization model}}.
\newblock \bibinfo{howpublished}{\url{https://techcommunity.microsoft.com/blog/windowsosplatform/openhcl-evolving-azure\%e2\%80\%99s-virtualization-model/4248345}}.
\newblock


\bibitem[Sabt et~al\mbox{.}(2015)]%
        {sabt2015trusted}
\bibfield{author}{\bibinfo{person}{Mohamed Sabt}, \bibinfo{person}{Mohammed Achemlal}, {and} \bibinfo{person}{Abdelmadjid Bouabdallah}.} \bibinfo{year}{2015}\natexlab{}.
\newblock \showarticletitle{Trusted execution environment: What it is, and what it is not}. In \bibinfo{booktitle}{\emph{2015 IEEE Trustcom/BigDataSE/Ispa}}, Vol.~\bibinfo{volume}{1}. IEEE, \bibinfo{pages}{57--64}.
\newblock


\bibitem[Scopelliti et~al\mbox{.}(2024)]%
        {scopelliti2024understanding}
\bibfield{author}{\bibinfo{person}{Gianluca Scopelliti}, \bibinfo{person}{Christoph Baumann}, {and} \bibinfo{person}{Jan~Tobias M{\"u}hlberg}.} \bibinfo{year}{2024}\natexlab{}.
\newblock \showarticletitle{Understanding Trust Relationships in Cloud-Based Confidential Computing}. In \bibinfo{booktitle}{\emph{2024 IEEE European Symposium on Security and Privacy Workshops (EuroS\&PW)}}. IEEE, \bibinfo{pages}{169--176}.
\newblock


\bibitem[Segarra et~al\mbox{.}(2019)]%
        {CCforMedTec}
\bibfield{author}{\bibinfo{person}{Carlos Segarra}, \bibinfo{person}{Ricard Delgado-Gonzalo}, \bibinfo{person}{Mathieu Lemay}, \bibinfo{person}{Pierre-Louis Aublin}, \bibinfo{person}{Peter Pietzuch}, {and} \bibinfo{person}{Valerio Schiavoni}.} \bibinfo{year}{2019}\natexlab{}.
\newblock \showarticletitle{Using {Trusted} {Execution} {Environments} for {Secure} {Stream} {Processing} of {Medical} {Data}}. In \bibinfo{booktitle}{\emph{Distributed {Applications} and {Interoperable} {Systems}}}, \bibfield{editor}{\bibinfo{person}{José Pereira} {and} \bibinfo{person}{Laura Ricci}} (Eds.). \bibinfo{publisher}{Springer International Publishing}, \bibinfo{address}{Cham}, \bibinfo{pages}{91--107}.
\newblock
\showISBNx{978-3-030-22496-7}
\urldef\tempurl%
\url{https://doi.org/10.1007/978-3-030-22496-7_6}
\showDOI{\tempurl}


\bibitem[{Synergy Research Group}(2024)]%
        {SynergyResearchGroup2025}
\bibfield{author}{\bibinfo{person}{{Synergy Research Group}}.} \bibinfo{year}{2024}\natexlab{}.
\newblock \bibinfo{title}{{Cloud Market Growth Surge Continues in Q3 {\textendash} Growth Rate Increases for the Fourth Consecutive Quarter {$\vert$} Synergy Research Group}}.
\newblock
\newblock
\urldef\tempurl%
\url{https://www.srgresearch.com/articles/cloud-market-growth-surge-continues-in-q3-growth-rate-increases-for-the-fourth-consecutive-quarter}
\showURL{%
\tempurl}
\newblock
\shownote{[Online; accessed 13. Jan. 2025]}.


\bibitem[{Technical University of Munich}(2025)]%
        {TUMSEVAnalysis}
\bibfield{author}{\bibinfo{person}{{Technical University of Munich}}.} \bibinfo{year}{2025}\natexlab{}.
\newblock \bibinfo{booktitle}{\emph{An Empirical Analysis of AMD SEV-SNP and Intel TDX}}.
\newblock \bibinfo{type}{Technical Report}. \bibinfo{institution}{TUM}.
\newblock
\urldef\tempurl%
\url{https://dse.in.tum.de/wp-content/uploads/2024/11/sigmetrics25summer-CVM-Explained.pdf}
\showURL{%
\tempurl}
\newblock
\shownote{Accessed on February 02, 2025}.


\bibitem[{Trusted Computing Group}(2019)]%
        {TPM2Spec}
\bibfield{author}{\bibinfo{person}{{Trusted Computing Group}}.} \bibinfo{year}{2019}\natexlab{}.
\newblock \bibinfo{booktitle}{\emph{Trusted Platform Module Library Specification, Family 2.0, Level 00, Revision 01.59}}.
\newblock \bibinfo{type}{Specification}. \bibinfo{institution}{Trusted Computing Group}.
\newblock
\urldef\tempurl%
\url{https://trustedcomputinggroup.org/resource/tpm-library-specification/}
\showURL{%
\tempurl}


\bibitem[Tsai et~al\mbox{.}(2017)]%
        {tsai17graphene-sgx}
\bibfield{author}{\bibinfo{person}{Chia-Che Tsai}, \bibinfo{person}{Donald~E. Porter}, {and} \bibinfo{person}{Mona Vij}.} \bibinfo{year}{2017}\natexlab{}.
\newblock \showarticletitle{Graphene-SGX: a practical library OS for unmodified applications on SGX}. In \bibinfo{booktitle}{\emph{Proceedings of the 2017 USENIX Conference on Usenix Annual Technical Conference}} (Santa Clara, CA, USA) \emph{(\bibinfo{series}{USENIX ATC '17})}. \bibinfo{publisher}{USENIX Association}, \bibinfo{address}{USA}, \bibinfo{pages}{645–658}.
\newblock
\showISBNx{9781931971386}


\bibitem[Ye et~al\mbox{.}(2022)]%
        {Hecate}
\bibfield{author}{\bibinfo{person}{Yuxin Ye}, \bibinfo{person}{Wenhao Shi}, \bibinfo{person}{Haibo Hu}, \bibinfo{person}{Zhengyu Gu}, \bibinfo{person}{Chengyu Qian}, \bibinfo{person}{Zhi Wang}, \bibinfo{person}{Xiang Qian}, \bibinfo{person}{Huibo Xue}, {and} \bibinfo{person}{Xinyu Zhang}.} \bibinfo{year}{2022}\natexlab{}.
\newblock \showarticletitle{Hecate: Lifting and Shifting On-Premises Workloads to an Untrusted Cloud}. In \bibinfo{booktitle}{\emph{Proceedings of the 2022 ACM SIGSAC Conference on Computer and Communications Security}}. \bibinfo{pages}{2775--2789}.
\newblock
\urldef\tempurl%
\url{https://doi.org/10.1145/3548606.3560592}
\showDOI{\tempurl}


\end{thebibliography}
\bibliographystyle{ACM-Reference-Format}

\end{document}